\documentclass[aps,onecolumn,showpacs]{revtex4}
\usepackage{float}
\usepackage{amsmath}
\usepackage{amssymb}
\usepackage{graphicx}
\usepackage{subfigure}

\begin{document}

\title{Single- and multi-peak solitons in two-component models of
metamaterials and photonic crystals}
\author{Peter Y. P. Chen$^{1}$, Boris A. Malomed$^{2}$}
\affiliation{$^{1}$School of Mechanical and Manufacturing Engineering, University of New
South Wales, Sydney 2052, Australia\\
$^{2}$Department of Physical Electronics, School of Electrical Engineering,
Faculty of Engineering, Tel Aviv University, Tel Aviv 69979, Israel}

\begin{abstract}
We report results of the study of solitons in a system of two
nonlinear-Schr\"{o}dinger (NLS) equations coupled by the XPM
interaction, which models the co-propagation of two waves in
metamaterials (MMs). The same model applies to photonic crystals
(PCs), as well as to ordinary optical fibers, close to the
zero-dispersion point. A peculiarity of the system is a small
positive or negative value of the relative group-velocity dispersion
(GVD) coefficient in one equation, assuming that the dispersion is
anomalous in the other. In contrast to earlier studied systems of
nonlinearly coupled NLS equations with equal GVD coefficients, which
generate only simple single-peak solitons, the present model gives
rise to families of solitons with complex shapes, which feature
extended oscillatory tails and/or a double-peak structure at the
center. Regions of existence are identified for single- and
double-peak bimodal solitons, demonstrating a broad bistability in
the system. Behind the existence border, they degenerate into
single-component solutions. Direct simulations demonstrate stability
of the solitons in the entire existence regions. Effects of the
group-velocity mismatch (GVM) and optical loss are considered too.
It is demonstrated that the solitons can be stabilized against the
GVM by means of the respective ``management" scheme. Under the
action of the loss, complex shapes of the solitons degenerate into
simple ones, but periodic compensation of the loss supports the
complexity.
\end{abstract}

\maketitle

\section{Introduction and the model}

Theoretical and experimental studies of various artificial optical media
have recently drawn a great deal of interest, see, e.g., Refs. \cite{art}
and references therein. Among these, a well-known class is formed by
metamaterials (MMs) based on the intrinsic periodic structures, built with
subwavelength characteristic scales, that feature a combination of negative
dielectric permittivity and magnetic permeability, and thus give rise to the
negative refractive index \cite{Pendry}. MMs promise a number of potential
applications impossible in ordinary optical media, such as superlensing \cite%
{superlense,Shalaev}. MMs are also described as ``left-handed"
waveguides, as they were originally predicted as those where the
wave vector of the electromagnetic wave is antiparallel to the usual
right-handed cross product of the electric and magnetic fields \cite%
{Veselago}. The intensive work has resulted in the creation of MMs featuring
the negative refraction at optical frequencies \cite{Shalaev}. Existing MMs
are usually assembled as periodic arrays of metallic split-ring resonators,
with various modifications of this basic setting.

Theoretical analysis of various nonlinear effects in models of MMs including
cubic \cite{theory}-\cite{Zhou} or quadratic \cite{Scalora,Haus} terms has
also drawn considerable attention. Some of these studies predict the
existence of solitons in models of the MM type \cite{solitons}. In
particular, a theoretical model for the XPM
(cross-phase-modulation)-mediated interaction of co-propagating waves,
carried through the MM at different frequencies, is similar to the
well-known two-wave model of ordinary media (where, in particular, it may
give rise to the modulational instability in the case of the normal
group-velocity dispersion (GVD) \cite{Agrawal}, and to domain-wall patterns
\cite{DW}, in the same case). However, an essential peculiarity of the MM
model is that the co-propagating waves may have widely different GVD
coefficients, even with opposite signs, corresponding to normal and
anomalous dispersion \cite{Zhou}. Another type of artificially built optical
media where this feature may be realized is represented by photonic crystals
(PCs) and PC fibers \cite{PCF,book}. In fact, these possibilities illustrate
the versatilely of artificial optical systems, which allow one to \textit{%
engineer} required properties, that may often be made very unusual.

The main subject of the present work is constructing families of stable
solitons within the framework of the system of XPM-coupled NLS
(nonlinear-Schr\"{o}dinger) equations describing the co-propagation of two
waves with local amplitudes $u(z,\tau )$ and $v\left( z,\tau \right) $,
carried by different frequencies:%
\begin{eqnarray}
iu_{z}+(1/2)u_{\tau \tau }+\left( |u|^{2}+2|v|^{2}\right) u &=&0,  \label{u}
\\
iv_{z}+icv_{\tau }+(1/2)D_{2}u_{\tau \tau }+\left( |v|^{2}+2|u|^{2}\right) v
&=&0.  \label{v}
\end{eqnarray}%
Here $z$ and $\tau $ are, as usual, the propagation distance and reduced
time, the GVD coefficient in Eq. (\ref{u}) is normalized to be $D_{1}\equiv 1
$ (which implies the anomalous sign of the GVD for wave $u$), $D_{2}$, that
may have either sign, is the relative dispersion for the second wave ($%
D_{2}<0$ implies that wave $v$ propagates with normal GVD), $c$ is the
walk-off parameter (alias the group-velocity mismatch, GVM, between the two
carrier frequencies), which is an essential ingredient of the respective MM
model \cite{SpecialTopics}, and it is assumed that effective
Kerr-nonlinearity coefficients are equal for both waves.

It is relevant to mention that the GVM between the fundamental-frequency and
second-harmonic waves also plays an important in role in the MM model with
the $\chi ^{(2)}$ nonlinearity. An interesting feature of that system is
that the phase locking of the second harmonic to the fundamental one, which
is strongly affected by the GVM, may effectively impose the action of the
negative refractive index on the second harmonic even in the situation when
it directly affects only the fundamental frequency \cite{Haus}.

Stationary solutions to Eqs. (\ref{u}) and (\ref{v}) with two independent
wavenumbers of the components, $K_{u,v}$, are looked for as
\begin{equation}
\left\{ u\left( z,\tau \right) ,v\left( z,\tau \right) \right\} =\exp \left(
iK_{u,v}z\right) \left\{ U(\tau ),V(\tau )\right\} .  \label{stationary}
\end{equation}%
In the case of $c=0$, the stationary waveforms $U,V$ may be assumed real,
obeying the equations following from the substitution of expressions (\ref%
{stationary}) into Eqs. (\ref{u}) and (\ref{v}):%
\begin{equation}
\begin{array}{c}
(1/2)U^{\prime \prime }+\left( U^{2}+2V^{2}\right) U=K_{u}U, \\
\left( D_{2}/2\right) U^{\prime \prime }+\left( V^{2}+2U^{2}\right) V=K_{v}V.%
\end{array}
\label{UV}
\end{equation}

Equations (\ref{u}) and (\ref{v}) were recently derived in the context of
MMs, neglecting the loss \cite{Zhou}. In fact, the presence of strong losses
is an inherent property of MMs \cite{loss}, and, moreover, it has been
rigorously proved that the negative refraction is not possible in a lossless
medium \cite{necessary-loss}. Nevertheless, schemes have been proposed to
essentially compensate the MM loss, using the matched impedance \cite%
{Shalaev,matched-impedance} or various gain mechanisms \cite{gain}. In any
case, it makes sense to study solitons in conservative models of the MM type
\cite{Zhou,no-loss}. In addition, the model based on Eqs. (\ref{u}) and (\ref%
{v}) may also be realized in PCs and PC fibers, where losses are present
too, but as a less severe problem \cite{PCF}. On the other hand, as concerns
solitons, a relevant approach may be to include a lossy MM sample, along
with an amplifier, as elements into an optical cavity, and predict cavity
solitons possible in such a setting \cite{cavity}.

As mentioned above, Eqs. (\ref{u}) and (\ref{v}) also apply to the
description of the co-propagation of XPM-interacting waves in usual optical
media, such as optical fibers \cite{Agrawal}, although in that case most
works have been dealing with the case of equal GVD coefficient, $D_{2}=1$.
Nevertheless, the case of $D_{2}<0$ is relevant too, if the two carrier
waves are chosen at wavelengths placed at opposite sides of the
zero-dispersion point of the optical fiber. Moreover, a matched pair of the
wavelengths may be chosen, so as to nullify the GVM between them ($c=0$).
The latter setting may find applications to fiber-optic telecommunications,
as the normal-GVD wave, $v$ (with $D_{2}<0$) may carry a stable periodically
modulated wave, that can be used as a support structure suppressing the
jitter in the data-carrying soliton stream launched in the anomalous-GVD
mode \cite{Ship}.

In previous studies of solitons supported by the XPM-coupled NLS equations,
families of bimodal (two-component) soliton solutions were studied in
detail, using both numerical methods \cite{numerical,variational} and the
variational approximation \cite{variational}, but only for the case of $%
D_{2}=1$ and $c=0$. In the general case, those solitons may feature
the ``elliptic polarization", i.e., an arbitrary ratio of
energies in the two components,%
\begin{equation}
\frac{E_{v}}{E_{u}}\equiv \frac{\int_{-\infty }^{+\infty }\left[ U(\tau )%
\right] ^{2}d\tau }{\int_{-\infty }^{+\infty }\left[ V(\tau )\right]
^{2}d\tau }  \label{E/E}
\end{equation}%
(in other words, the solitons may exist with arbitrary positive values of
ratio $R\equiv K_{v}/K_{u}$ of the two wavenumbers), but they were simplest
fundamental single-humped solitons. The coupled-NLS system with $D_{2}=1$
may have multi-hump soliton solutions too, but they are expected to be
unstable \cite{Yang,aboutMoti2}. In this paper, we aim to demonstrate that,
for values $D_{2}<1$, including small \emph{negative} values of the relative
GVD coefficient (i.e., the case of weak \emph{normal} GVD in the $v$%
-component), Eqs. (\ref{u}), (\ref{v}) and (\ref{UV}) give rise to
very different families of stable single- and multi-humped solitons,
which, in particular, entails \emph{bistability} of the soliton
states in a broad range of parameters. A characteristic feature of
the solitons revealed by the analysis at small positive and negative
values of $D_{2}$ is the existence of weakly localized oscillating
\textit{tails} attached to the main ``body" of the bimodal soliton.

It is relevant to mention that multi-hump solitons in multi-mode optical
systems were first predicted \cite{Moti1} and experimentally observed \cite%
{Moti2} in photorefractive media, with the saturable nonlinearity, which
induces the XPM interactions between different guided modes. The stability
of such multi-hump states was later demonstrated in a rigorous form \cite%
{aboutMoti1,aboutMoti2}. These studies were also extended into the
two-dimensional setting \cite{Moti2D}.

Basic results concerning the existence and stability (including the
bistability) of the solitons in the present model without the GVM
($c=0$) are summarized in Section 2. A the end of this section, we
consider the action of the loss on solitons, and periodic
compensation of the loss. Section 3 is dealing with an approach that
allows one to stabilize solitons against the GVM by means of the
``mismatch-management" technique: we consider a model with
coefficient $c$ which periodically jumps between large positive and
negative values, making the average GVM equal to zero. Physically,
this implies the use of a layered medium composed of segments with
opposite values of the GVM. In fact, the GVM is known to be a
serious issue in a different context, \textit{viz}., an obstacle to
the creation of temporal solitons in optical media with the
quadratic ($\chi ^{(2)}$) nonlinearity \cite{chi2,Kale}. In that
context, the GVM-management technique was recently elaborated in a
theoretical form \cite{we1}. Actually, it resembles the previously
proposed ``tandem" schemes, which provide for an effective
compensation of the GVM by way of a
periodic alternation of linear and $\chi ^{(2)}$-nonlinear layers \cite%
{tandem}. The paper is concluded by Section 4.

\section{Solitons and their stability regions}

\subsection{The numerical method}

The underlying equations (\ref{u}), (\ref{v}), as well as their stationary
counterparts, Eqs. (\ref{UV}), were solved in the numerical form by means of
a pseudospectral method, which was adjusted to the present model, following
the lines of works \cite{pseudo} (the application of this method to soliton
solutions of equations of the NLS type was recently elaborated in Ref. \cite%
{pseudo-soliton}). The integration domain of variable $\tau $, of width $%
T=40 $, was sufficient to completely study the shape and dynamics of
individual solitons, including those which feature extended ``tails"
(simulations of interactions between solitons might require using a
larger domain). The domain was divided into $N$ sub-domains,
the fields in each one being approximated by truncated power expansions,%
\begin{equation}
\left\{ u_{j}\left( \tau ,z\right) ,v_{j}\left( \tau ,z\right) \right\}
=\sum_{k=1}^{M}\left\{ u_{jk}\left( z\right) ,v_{jk}\left( z\right) \right\}
\left( \frac{\tau -\tau _{j}^{(0)}}{\Delta \tau }\right) ^{k},
\label{expansion}
\end{equation}%
$j=1,...,N$, where $\tau _{j}^{(0)}$ is the midpoint of the sub-domain, and $%
\Delta \tau \equiv T/\left( 2N\right) $ its half-width. The substitution of
expressions (\ref{expansion}) into Eqs. (\ref{u}), (\ref{v}), truncation of
ensuing expansions, and the use of the collocation at Chebyshev points and
the continuity conditions for the functions and their first derivatives at
boundaries between the sub-domains lead to a system of ordinary differential
equations (ODEs) for amplitudes $u_{jk}\left( z\right) $ and $v_{jk}\left(
z\right) $, which were numerically solved by means of an unconditionally
stable implicit Crank-Nicolson scheme. Accordingly, stationary solitons were
looked for as $z$-independent solutions of the ODE system, reducing it to a
system of algebraic equations. The latter one was solved using the Newton's
method. Sufficient accuracy of the results could be usually achieved with $%
M=N=8$.

As we demonstrate below, the choice of the pseudospectral method is
essential for obtaining reliable numerical solutions, while a more common
approach, based on the split-step fast-Fourier-transform (SSFFT) scheme
(see, e.g., \cite{split-step}), may encounter problems in the most essential
case considered in this work, \textit{viz}., simulations of solitons with
well-pronounced ``tails". Further details of the numerical procedure, which
may be useful for other applications too, will be reported elsewhere.

\subsection{Two types of solitons}

\begin{figure}[tbp]
\subfigure[]{\includegraphics[width=3in]{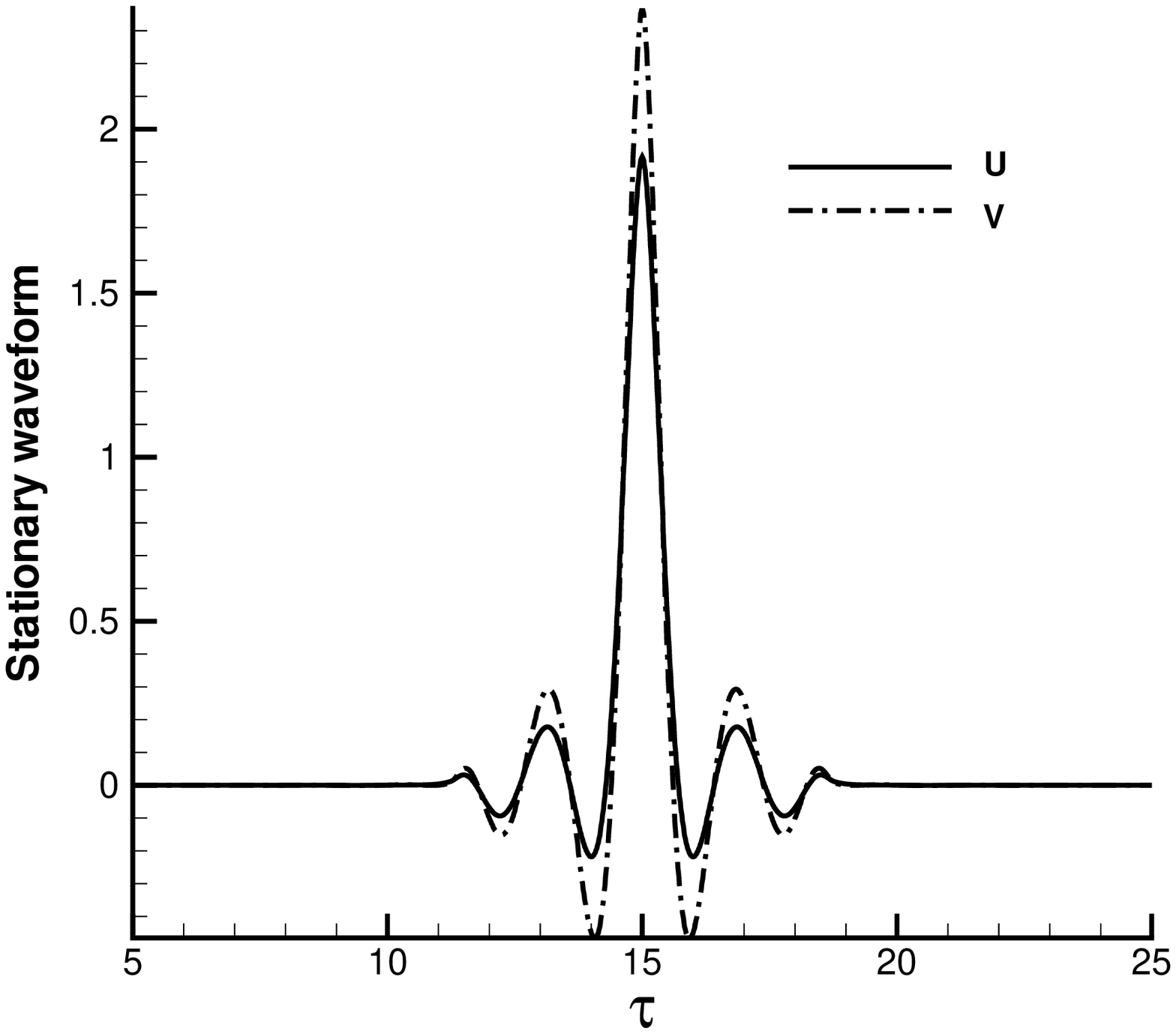}}\subfigure[]{%
\includegraphics[width=3in]{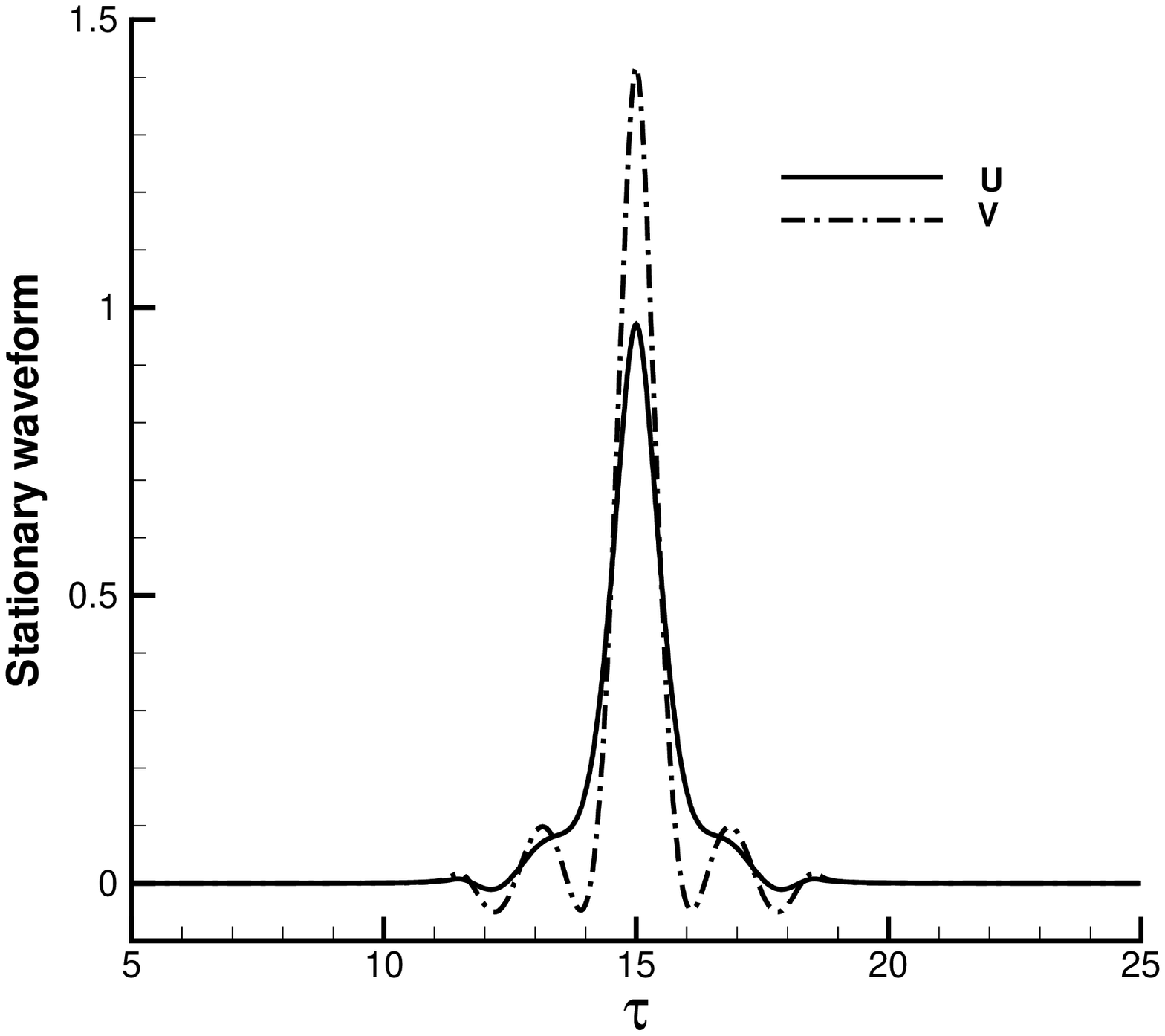}}
\caption{Examples of stable bimodal solitons with the single central peak.
The total energy of the soliton, defined as per Eq. (\protect\ref{E}), is $%
E=5$ in (a) and $E=2$ in (b), respectively. The other parameters are $%
D_{2}=0.37,$ $c=0$, and the ratio of the wavenumbers of the two components
is $R\equiv K_{v}/K_{u}=1.13$ in both cases.}
\label{fig1}
\end{figure}

\begin{figure}[tbp]
\includegraphics[width=4in]{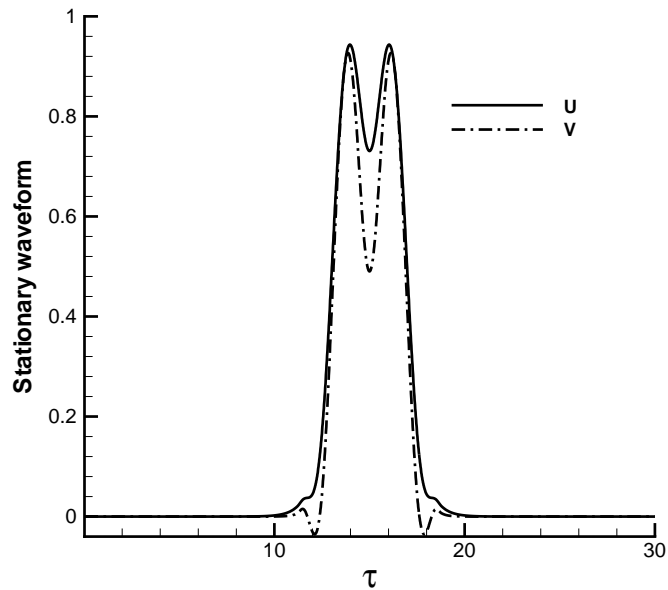}
\caption{An example of a stable double-hump soliton found at the same values
of parameters as in Fig. \protect\ref{fig1}(a), including the total energy
and the wavenumber ratio. }
\label{fig2}
\end{figure}

The numerical solution produces stationary solitons of two essentially
different types, \textit{viz}., ones with a single peak (\textit{hump}) at
the center of the soliton, as shown in Fig. \ref{fig1}, and double-hump
solitons with two separated main symmetric peaks, see an example in Fig. \ref%
{fig2}. These figures also illustrate the \emph{bistability} supported by
the present model, as the single- and double-hump solitons displayed in
Figs. \ref{fig1}(a) and \ref{fig2} exist at exactly the same value ($E=5)$
of the total energy,%
\begin{equation}
E=\int_{-\infty }^{+\infty }\left[ \left\vert u(\tau )\right\vert
^{2}+\left\vert v(\tau )\right\vert ^{2}\right] d\tau  \label{E}
\end{equation}%
(cf. Eq. (\ref{E/E})), and share a common value ($R=1.13$) of the wavenumber
ratio that was defined above, $R\equiv K_{v}/K_{u}$.

\begin{figure}[tbp]
\subfigure[]{\includegraphics[width=3in]{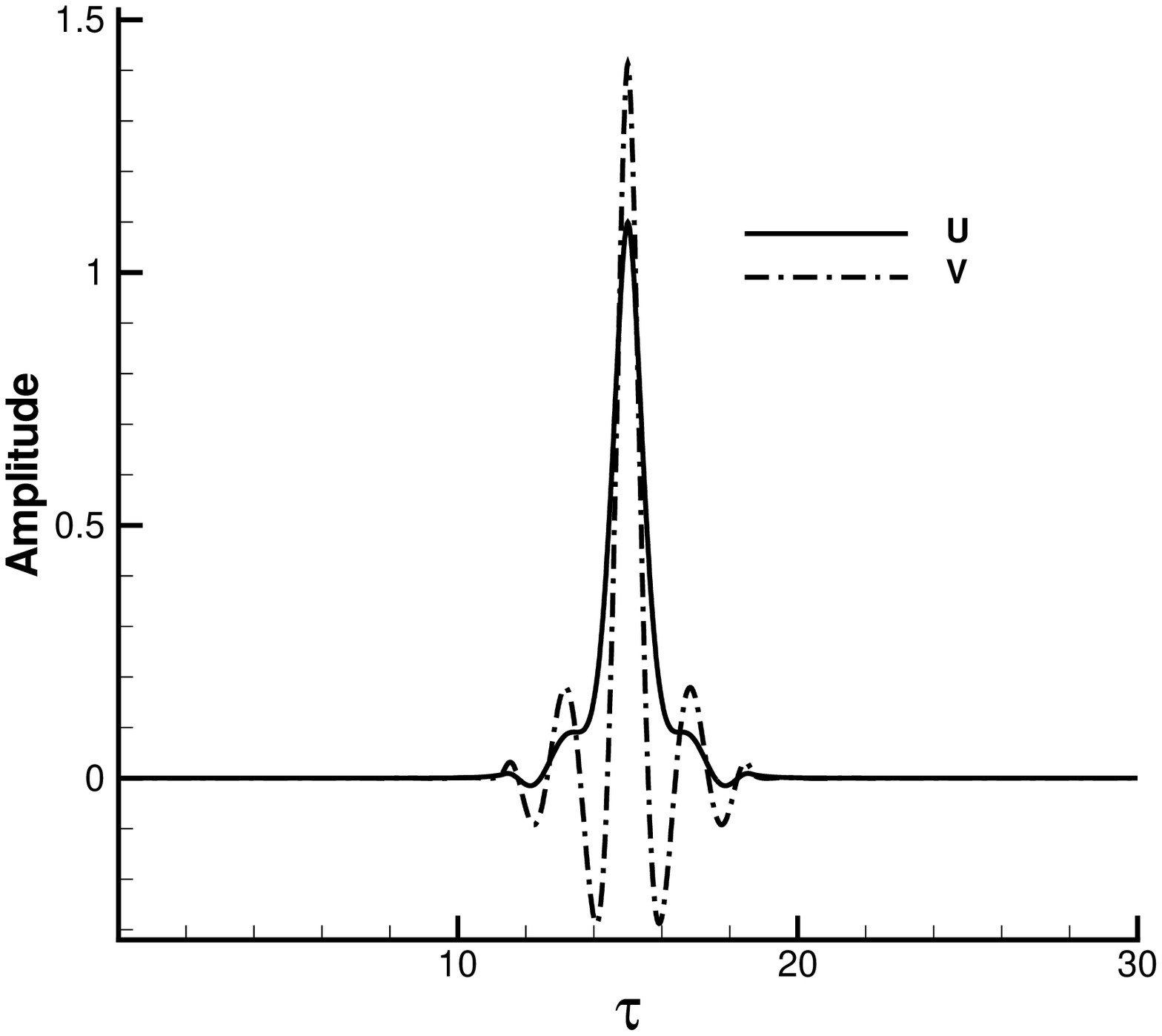}}\subfigure[]{%
\includegraphics[width=3in]{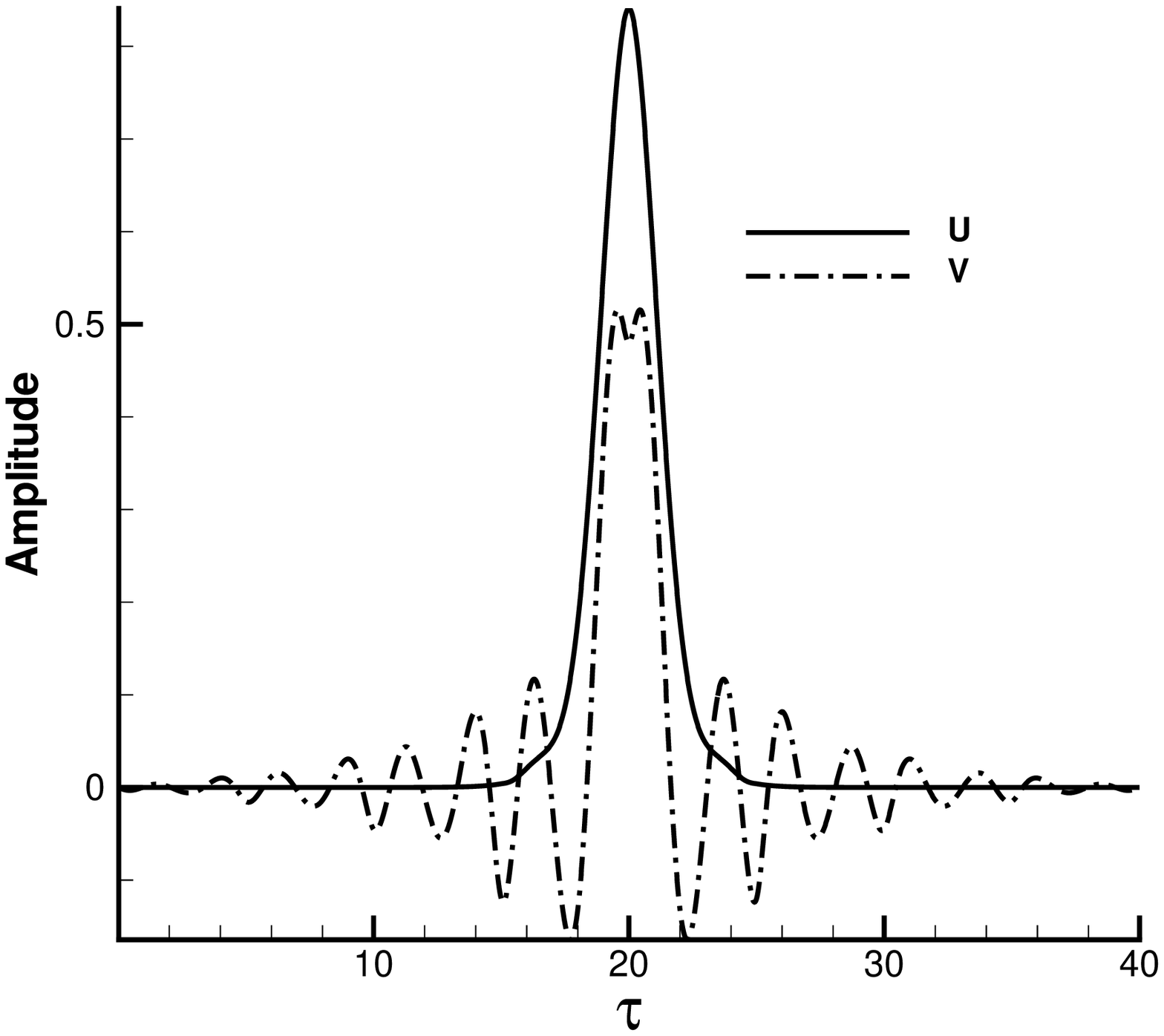}}
\caption{Examples of stable stationary solitons demonstrating the
growth of the ``tails" with the decrease of the GVD coefficient in
the
second component, $D_{2}$, at $c=0$, $E=2,$ $R=2.$ In (a) and (b), $%
D_{2}=0.1 $ and $D_{2}=-0.3$, respectively.}
\label{fig3}
\end{figure}

Further, Fig. \ref{fig3} shows that, in either case (single- or double-peak
structure), the solitons develop slowly decaying oscillatory tails as the
relative GVD coefficient in Eq. (\ref{v}), $D_{2}$, takes smaller values. In
fact, the problem posed by strong tails in bimodal systems with (slightly)
normal GVD in one component is well known in the studies of optical models
with the $\chi ^{(2)}$ nonlinearity, as in the cases of practical interest $%
\chi ^{(2)}$ crystals feature normal GVD at the second harmonic \cite{chi2}.
While, strictly speaking, the tail in the normal-GVD component cannot vanish
at $\tau \rightarrow \pm \infty $, it has been concluded that one can
identify a well-defined region in the respective parameter space, including
a finite interval of negative values of $D_{2}$ (in the present notation),
where the amplitude of the normal-GVD field assumes so small values at $\tau
\rightarrow \infty $ that the nonvanishing portion of the tail is invisible
and may be regarded as nonexistent, for all practical purposes. A transition
from this case to the situation with a conspicuous long-range tail is sharp,
and may be easily identified \cite{Isaac,Kale,we1}. In the present model,
the tail is also, strictly speaking, nonvanishing in the case shown in Fig. %
\ref{fig3}(b), which pertains to $D_{2}=-0.3$. However, the figure
demonstrates that, in practical terms, the tail is absent at large values of
$|\tau |$.

Note that Fig. \ref{fig3}(b), which features, in addition to the
well-developed tails, a slightly split peak in the $v$ component, also
illustrates the transition between the single-hump and dual-hump types of
the solitonic shape. Particular values of the scaled parameters for which
Figs. \ref{fig1}-\ref{fig3} have been plotted (as indicated in captions to
the figures) were taken pursuant to typical values of physical parameters in
the lossless Drude model, which was used in Ref. \cite{Zhou} to derive the
coupled system of Eqs. (\ref{u}) and (\ref{v}) for the bimodal propagation
of electromagnetic waves in the region of the negative refractive index.

Unlike the situation for simple fundamental solitons which was
considered for $D_{2}=1$ \cite{variational}, the variational
approximation (VA)\ for solitons is not helpful in the present case,
as one would have to use a complex ansatz approximating the
oscillatory tails, as well the possibility of having the split peak
at the center of the soliton. The use of the VA for the description
of ``tailed" gap solitons in the model with the self-defocusing
nonlinearity and periodic potential (optical lattice) \cite{Arik},
as well as for the transition from unsplit to split solitons
profiles in two-component models \cite{Sadhan}, has demonstrated
that this may be possible in a limited range of parameters, but the
necessary calculations are quite cumbersome. Actually, direct
numerical solutions may be easier to obtain in this case than their
VA counterparts.

The knowledge of the exact shape of the solitons may be essential for
experiments, as the available size of MM samples, as well as the size of
PCs, may be quite small, hence the transmission distance in the sample may
not be long enough to allow self-shaping of the input signal into the
soliton. Therefore, to observe and use the soliton propagation, it may be
necessary to prepare an input pulse (in the case of the MM) or spatial beam
(to be coupled into a PC) whose shape is as close as possible to that of the
stable soliton, thus minimizing detrimental shape oscillations in the course
of the transmission.

\subsection{Stability regions}

The stability of all the stationary solitons was tested by means of direct
simulations of the propagation within the framework of Eqs. (\ref{u}) and (%
\ref{v}), adding small perturbations to the initial profile. As a
result, it has been concluded that all the solitons which could be
found in the stationary form, as reported above, are stable,
including those with extended tails, which were found at small
positive \emph{and negative} values of $D_{2}$. Moreover, it has
been concluded that the propagation distance necessary for the
self-healing of weakly perturbed solitons and their relaxation back
to the unperturbed shape was always essentially smaller than $z=10$
(in the present notation). Examples of the stable evolution of
solitons with complex shapes, which feature strong tails or the
split-peak structure, are displayed below in Fig. \ref{fig8}, see
also other relevant examples in Figs. \ref{fig5} and \ref{fig10}(a).

\begin{figure}[tbp]
\subfigure[]{\includegraphics[width=3in]{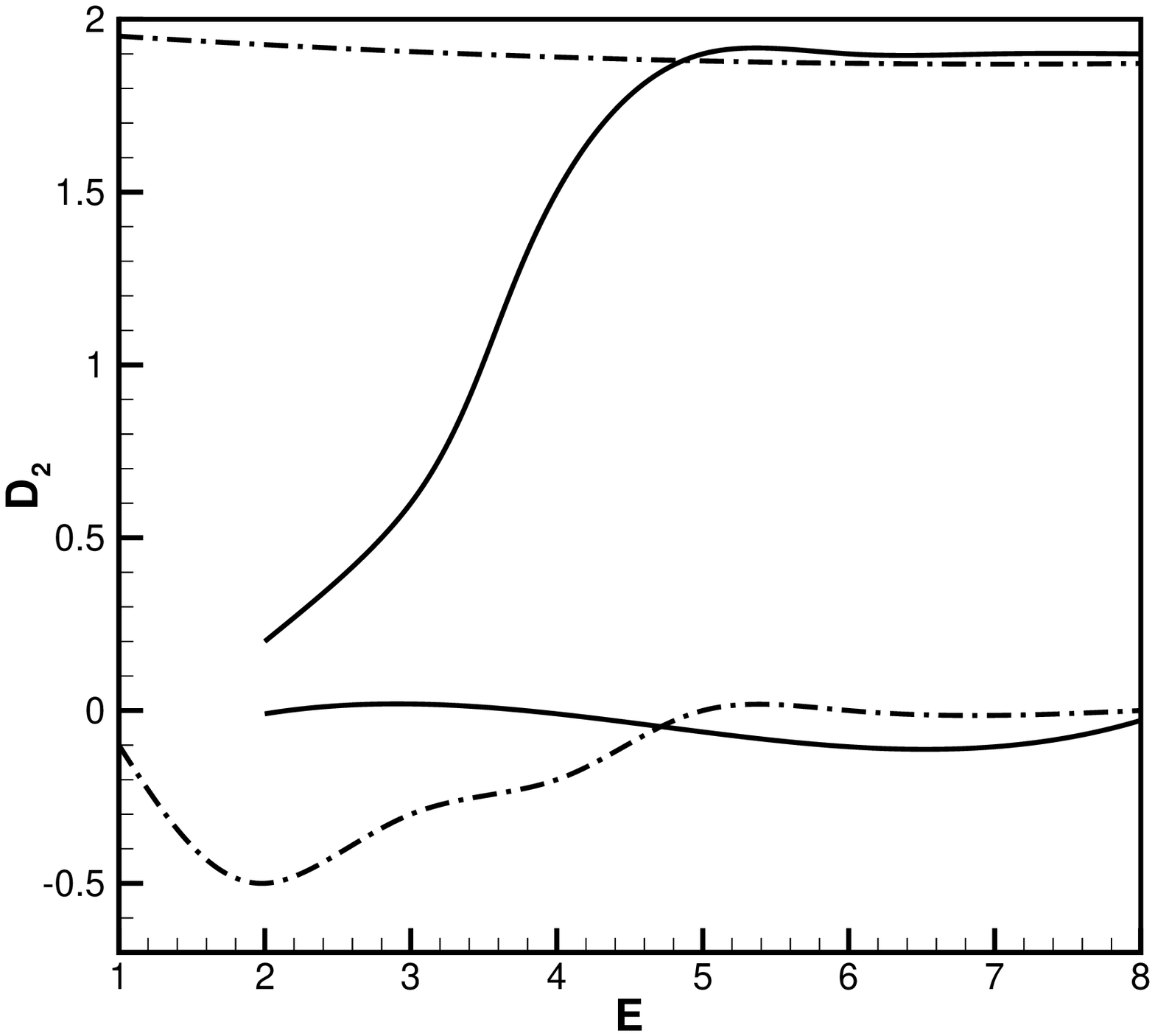}}\subfigure[]{%
\includegraphics[width=3in]{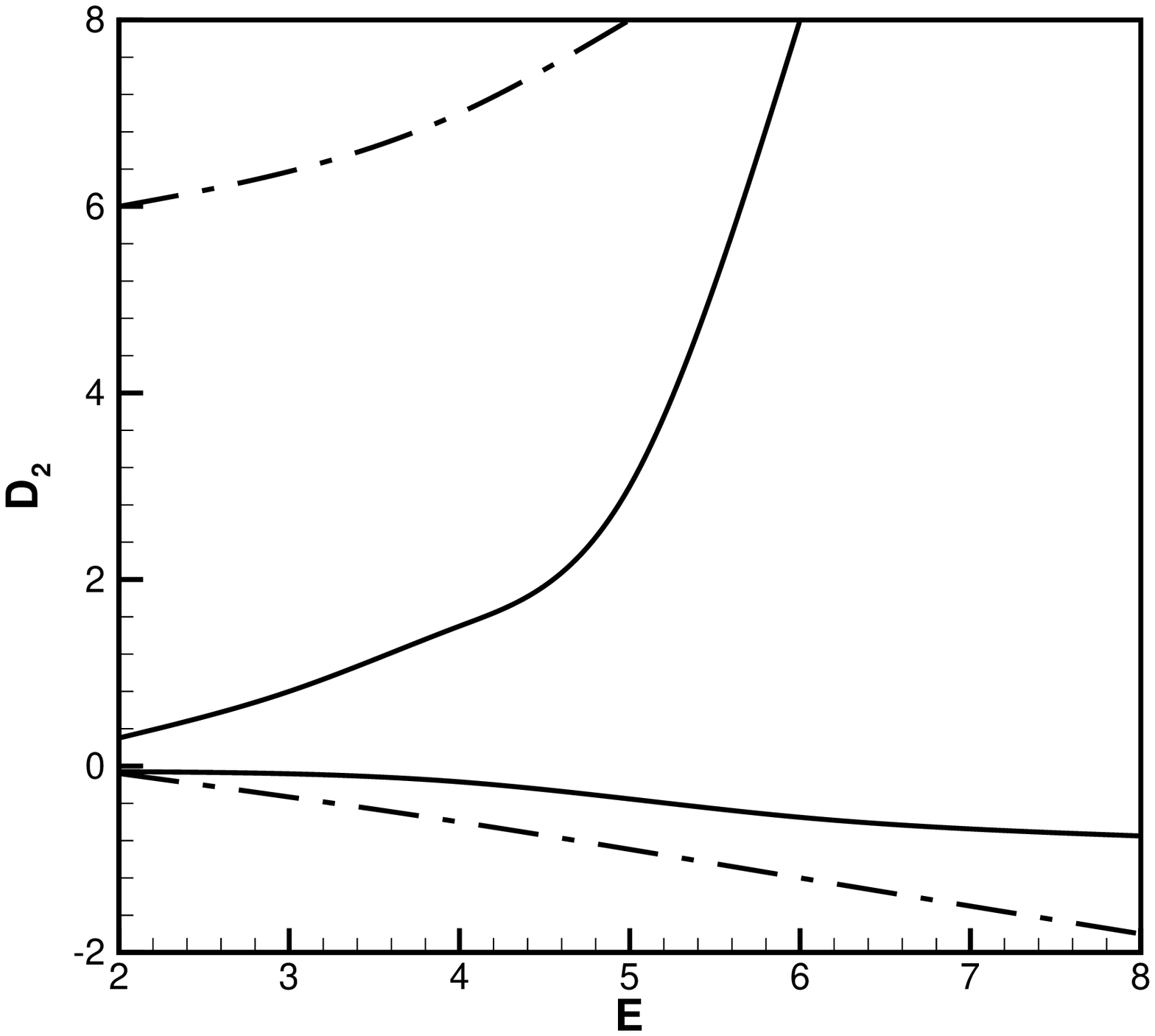}}
\caption{Borders of the regions of the existence (and, presumably,
stability) for the single-peak and double-peak solitons, in the model with $%
c=0$ (zero group-velocity mismatch) are shown by dashed-dotted and
continuous lines, respectively. Panels (a) and (b) correspond to
fixed values of the wavenumber ratio: $R\equiv K_{v}/K_{u}=2$ in
(a), and $R=1.2$ in (b). The overlap of the stability areas for both
species of the solitons demonstrates the bistability of the system.
In parts of the existence regions located at $D_{2}<0$ the
$v$-component of the solutions, strictly speaking, cannot take a
localized form. In fact, the bottom existence border separates the
solutions which are ``practically localized" and those with
conspicuous tails, that do not vanish in the entire integration
domain. For that reason, extension of the domain may lead to slight
expansion of the existence region. There is no right border for the
existence of the solitons, as they persist to indefinitely large
values of the energy.} \label{fig4}
\end{figure}

\begin{figure}[tbp]
\includegraphics[width=4in]{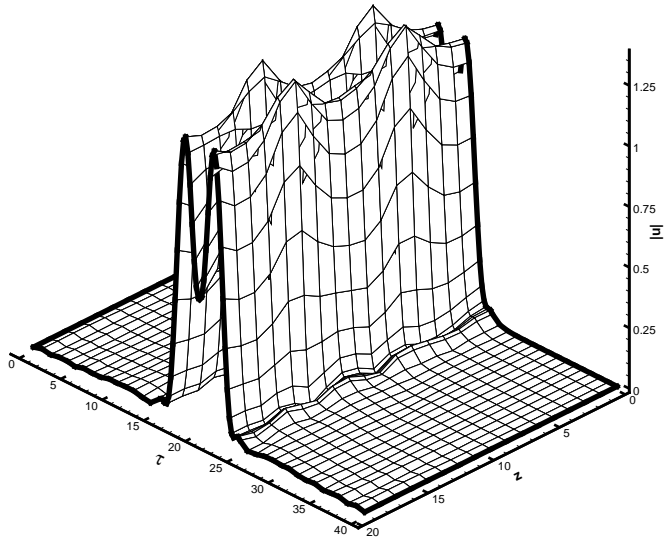}
\caption{The perturbed evolution of the soliton for $D_{2}=1$, when
its energy, $E=5$, was suddenly increased by $5\%$. The wavenumber
ratio \ of the unperturbed soliton is $R=2$.} \label{fig5}
\end{figure}

Thus, the stability region for the bimodal solitons is supposed to be
identical to their existence area. For fixed material constants, $D_{2}$ and
$c$ (actually, in this section we consider the case of $c=0$), soliton
families depend on two intrinsic parameters, that may be identified as their
wavenumbers $K_{u}$ and $K_{v}$, or, more conveniently, as the total energy,
$E$, and the wavenumber ratio, $R=K_{v}/K_{u}$. An adequate rendition of the
existence areas for the solitons of both types that were defined above,
i.e., single- and double-hump ones, is provided by plotting the areas in the
plane of $\left( E,D_{2}\right) $ at fixed values of $R$. In Fig. \ref{fig4}%
, we display overlapping existence regions of both soliton species for two
characteristic values of the wavenumber ratio, $R=2$ and $1.2$. The large
region of the bistability is evident in this figure.

As one approaches the upper existence border for the solitons of either
type, increasing $D_{2}$ at a fixed value of total energy $E$, the energy
ratio, $E_{v}/E_{u}$ (see Eq. (\ref{E/E})), decreases, vanishing at the
upper border. Above this border, only single-component solitons exist, with $%
v=0$ (of course, these are regular single-peak NLS solitons). The opposite
trend is observed with the decrease of $D_{2}$ at fixed $E$: as one
approaches the lower existence border, the $u$-component of the soliton
tends to vanish, along with its energy share, while the $v$-component
develops strong tails, especially in parts of the existence region located
at $D_{2}<0$. It is easy to explain these trends, looking at the Hamiltonian
of the present model (with $c=0$),%
\begin{equation}
H=\int_{-\infty }^{+\infty }\left[ (1/2)\left( \left\vert u_{\tau
}\right\vert ^{2}+D_{2}\left\vert v_{\tau }\right\vert
^{2}+|u|^{4}+|v|^{4}\right) +2|uv|^{2}\right] d\tau .  \label{H}
\end{equation}%
Indeed, the minimization of the gradient part of the Hamiltonian at fixed $E$
is obviously favored by the attenuation of the $v$-component for large $%
D_{2} $, and by the decay of the $u$-component for small $D_{2}$. Another
feature observed in Fig. \ref{fig4} is the shrinkage of the existence
regions for the double-hump solitons as the wavenumber ratio approaches $R=1$
(as suggested by the comparison of the panels appertaining to $R=2$ and $%
R=1.2$). At $R=1$ (equal wavenumbers in the two component), the system gives
rise to single-hump solitons only.

A mathematically rigorous approach to the stability analysis for the
solitons is based on the computation of the corresponding eigenvalues, using
equations for perturbation modes linearized around the stationary soliton
solutions. In particular, in Ref. \cite{aboutMoti2} is was concluded that
all multi-hump solitons may be unstable, in this sense, in the case of $%
D_{2}=1$. On the other hand, our direct simulations of the evolution
of double-hump solitons did not reveal any tangible instability at
$D_{2}=1$ (note that line $D_{2}=1$ goes across the solitons'
existence regions in Fig. \ref{fig4}): weakly perturbed solitons
remain apparently stable in this case, while a strong perturbation
excites intrinsic vibrations in the soliton, but does not destroy
it, see an example in Fig. \ref{fig5}. A possible explanation to
these findings (thanks to which the solitons are classified here as
stable also at $D_{2}=1$) is that the respective instability
eigenvalues may be very small.

\subsection{The split-step method and ``tailed" solitons}

\begin{figure}[tbp]
\includegraphics[width=4in]{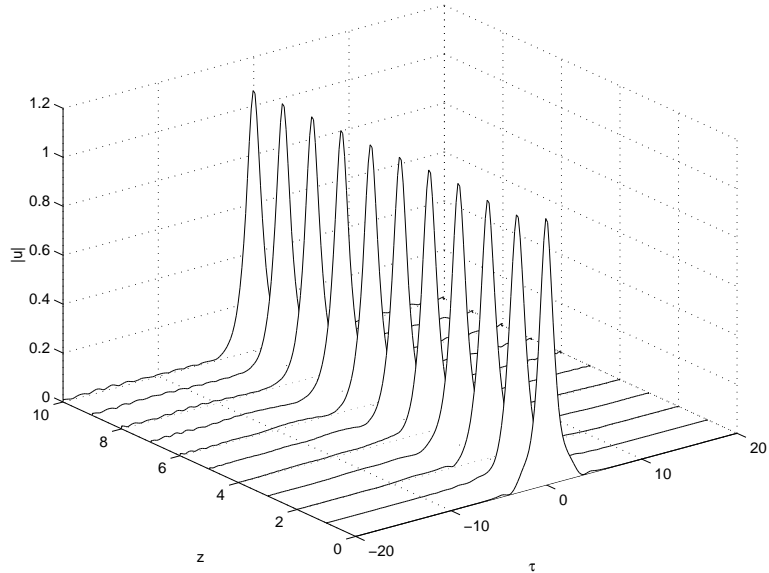}
\caption{Results of the application of the split-step
fast-Fourier-transform (SSFFT) method to the simulation of the
evolution of a single-peak soliton without well-pronounced ``tails",
for $D_{2}=0.3$ and $E=2$, $R\equiv E_{v}/E_{u}=2$. In this case,
the SSFFT algorithm is relatively stable.} \label{fig6}
\end{figure}

\begin{figure}[tbp]
\subfigure[]{\includegraphics[width=3in]{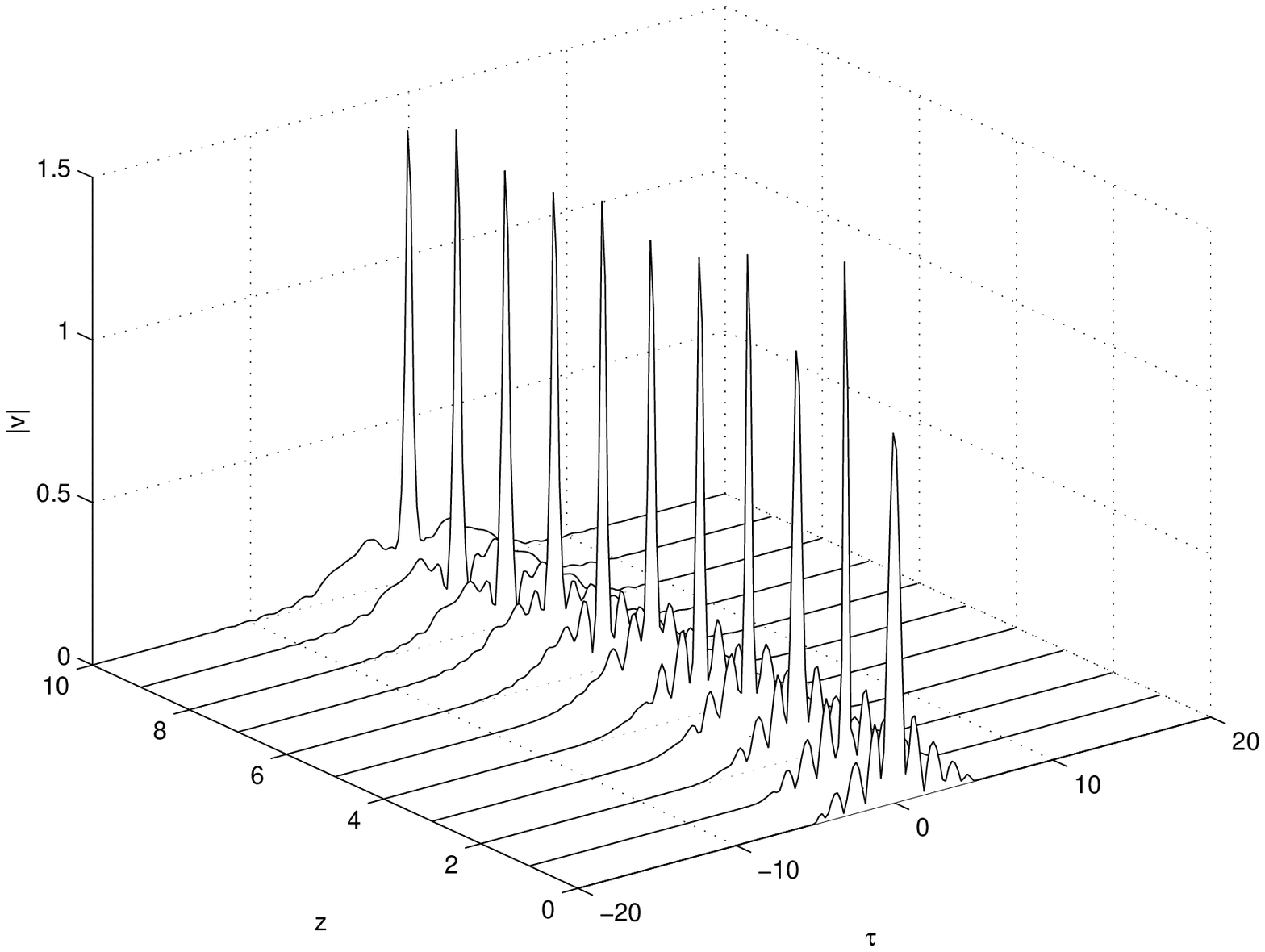}}\subfigure[]{%
\includegraphics[width=3in]{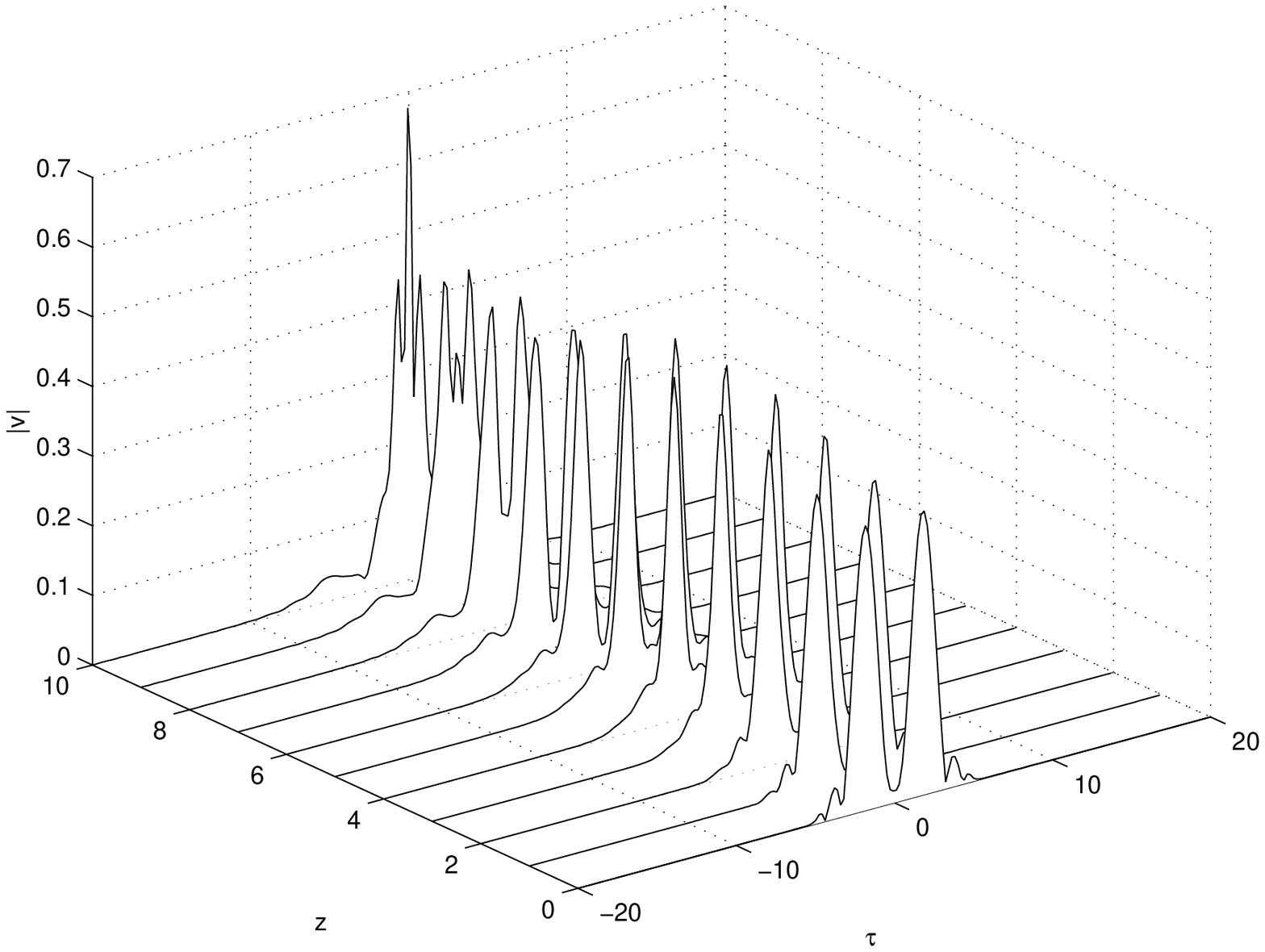}}
\caption{Unstable operation of the SSFFT algorithm simulating the evolution
of a single-peak soliton with conspicuous tails (a), and a double-peak
soliton (b). In both cases, the relative-GVD coefficient in Eq. (\protect\ref%
{v}) is $D_{2}=0.1$, and the initial solitons have $E=2$ and
$R\equiv K_{u}/K_{u}=2$.} \label{fig7}
\end{figure}

As said above, the application of the pseudospectral simulation
algorithm readily corroborates the stability of all solitons that
can be found in the stationary form, including those which feature
conspicuous ``tails" (see, e.g., Fig. \ref{fig3}). On the other
hand, it is relevant to mention that the popular
split-step-fast-Fourier-transform (SSFFT)\ method may produce
numerical problems in the case of strong ``tails". An example of
this is shown in Figs. \ref{fig6} and \ref{fig7}, which demonstrate
the difference in the numerical stability of the SSFFT algorithm in
the absence of well-pronounced tails, and in cases when they are
present.

We stress that the evolution of the solitons was simulated by means
of the SSFFT without adding any initial perturbations, i.e., Figs.
\ref{fig7}(a) and (b) display results generated by an intrinsic
instability of the algorithm when it is applied to the tailed and
double-hump solitons. The SSFFT was realized, in these examples,
covering the integration domain by 2048 points.

In fact, the solitons which were used as initial conditions in Figs. \ref%
{fig6} and \ref{fig7} are fully stable. To demonstrate this, in Fig. \ref%
{fig8} we display simulations of the same pair of solitons as in Fig. \ref%
{fig7}, but performed by means of the pseudospectral algorithm which
was outlined above.
\begin{figure}[tbp]
\subfigure[]{\includegraphics[width=3in]{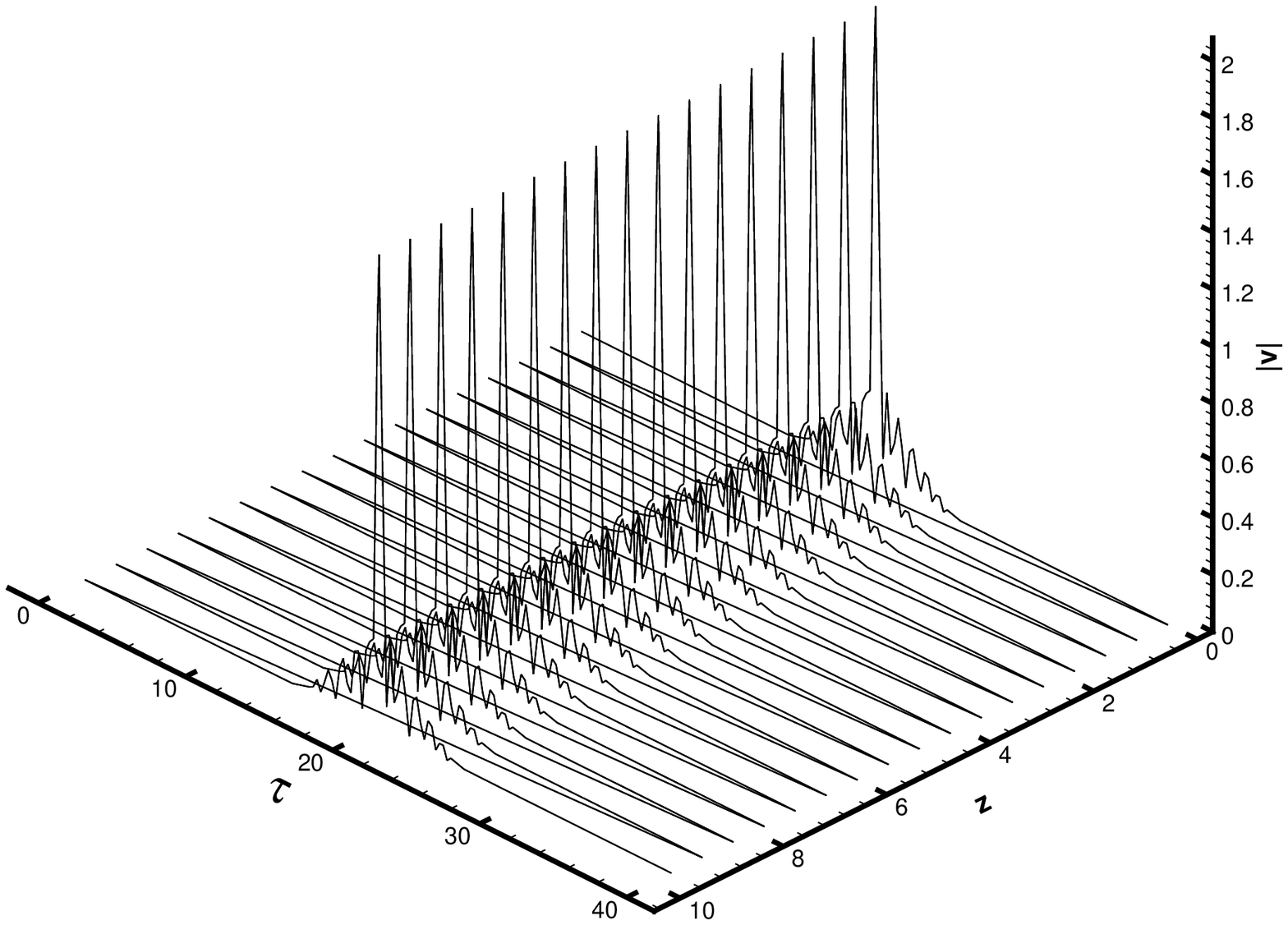}}\subfigure[]{%
\includegraphics[width=3in]{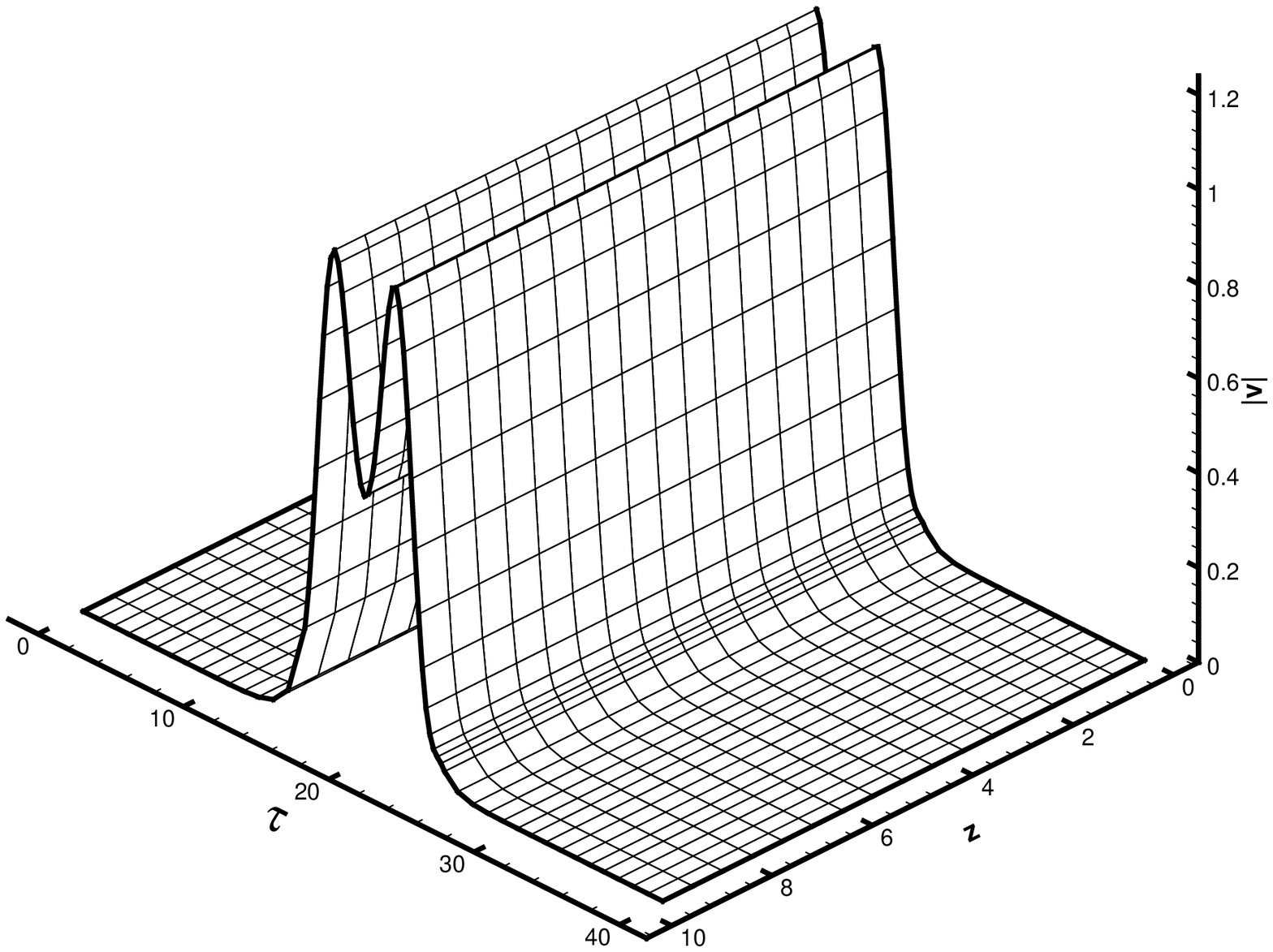}}
\caption{Simulations of the same initial solitons as in Fig. \protect\ref%
{fig7}, but carried out by means of the stable pseudospectral
method.} \label{fig8}
\end{figure}

\subsection{Effects of the loss and its compensation}

As said above, dissipative effects may be negligible under experimentally
relevant conditions in some types of artificial optical media, such as PCs
and PC fibers, but they constitute an essential ingredient in the
description of the light transmission through MMs \cite{loss,necessary-loss}%
. In the simplest approximation, the loss is taken into account by means of
a positive dissipative coefficient, $\alpha $, added to Eqs. (\ref{u}) and (%
\ref{v}):
\begin{eqnarray}
iu_{z}+(1/2)u_{\tau \tau }+\left( |u|^{2}+2|v|^{2}\right) u &=&-i\alpha u,
\label{au} \\
iv_{z}+icv_{\tau }+(1/2)D_{2}u_{\tau \tau }+\left( |v|^{2}+2|u|^{2}\right) v
&=&-i\alpha v.  \label{av}
\end{eqnarray}

\begin{figure}[tbp]
\subfigure[]{\includegraphics[width=3in]{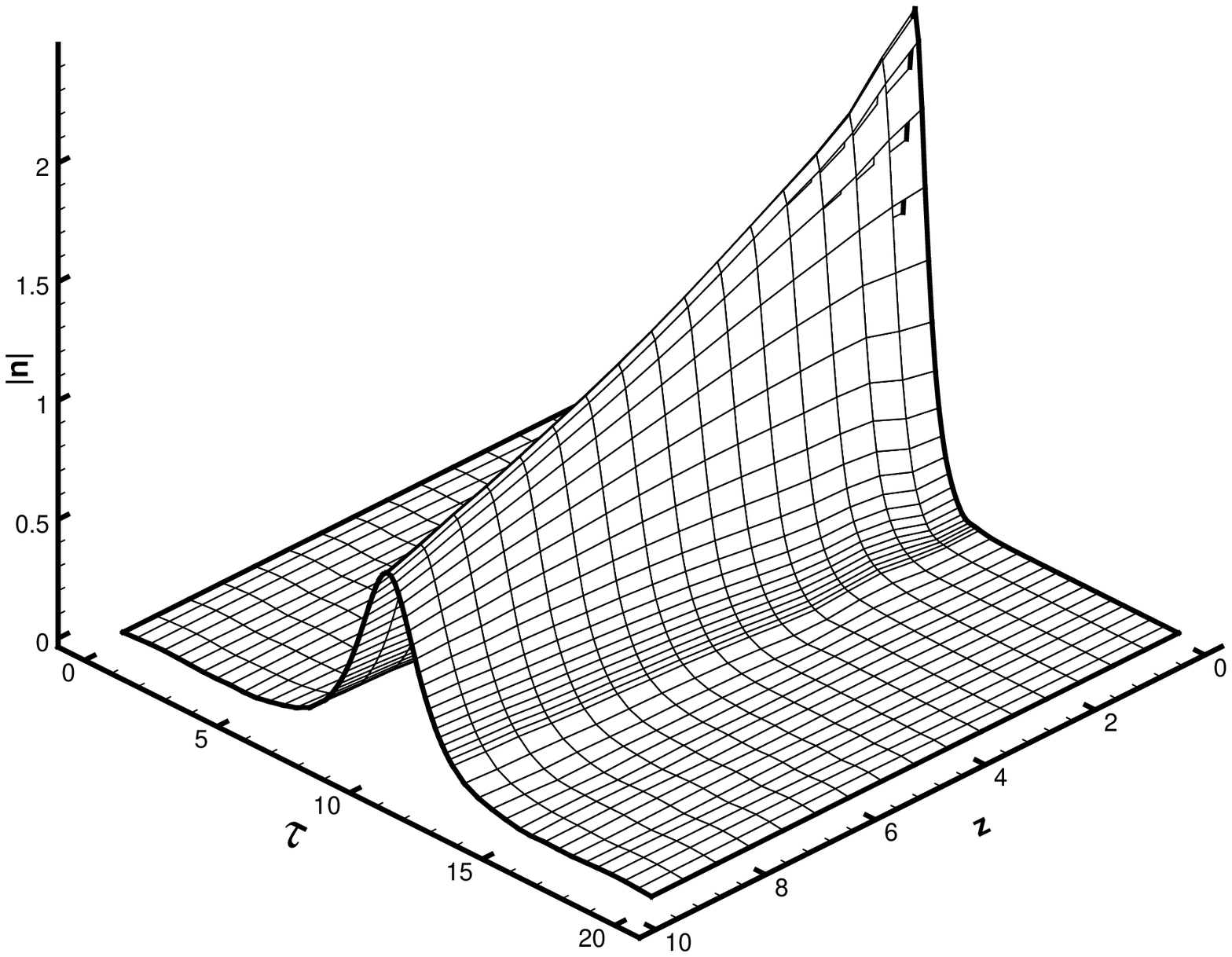}}\subfigure[]{%
\includegraphics[width=3in]{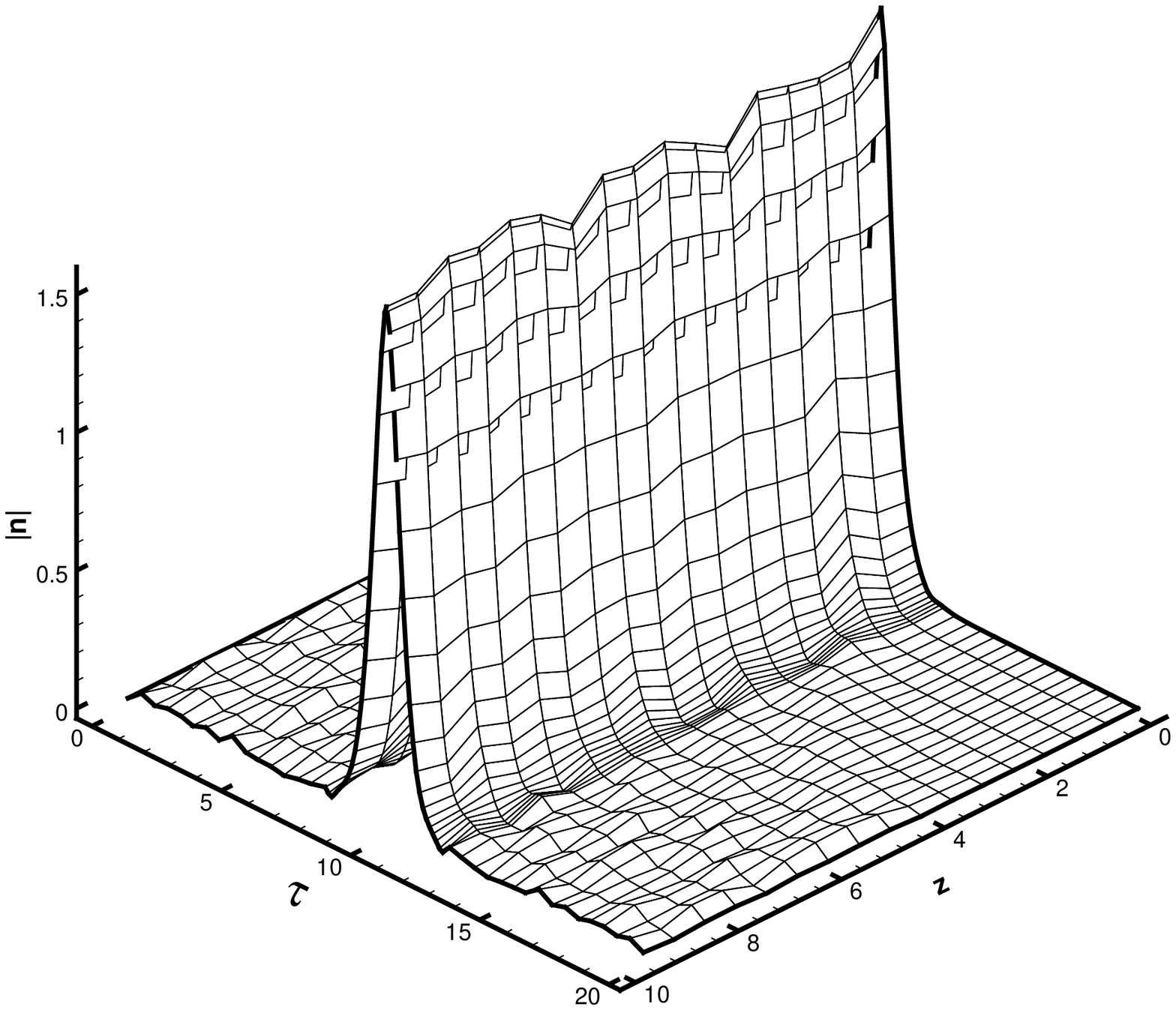}}
\caption{(a) Evolution of a typical soliton, with an originally
``tailed" shape, initial total energy $E_{0}=5$, and $%
R_{0}\equiv K_{v}/K_{u}=1.13$, under the action of the loss with $\protect%
\alpha =0.05$ and $D_{2}=0.37$ in Eqs. (\protect\ref{au}) and (\protect\ref%
{av}). (b) The same, but with the compensation of the loss by means
of the instantaneous linear amplification, applied with period
$\Delta z=1$.} \label{fig9}
\end{figure}

Obviously, the action of the loss leads to the decrease of the
soliton's energy. As seen in Fig. \ref{fig4}, the gradual decrease
of the total energy should drive the solitons in the direction of
the transition from the double-hump shape to the single-hump one,
i.e., the loss is expected to make the solitons' shape ``more
primitive". This expectation is corroborated by a typical example of
the evolution of the soliton under the action of the loss displayed
in Fig. \ref{fig9}(a): under the action of the loss, the soliton
gradually decays, transiting from the ``tailed" shape to a tailless
one.

As mentioned above, a number of works considered possibilities to
periodically compensate the loss in MMs by an optical gain \cite%
{Shalaev,matched-impedance,gain}. In Fig. \ref{fig9} we display an
example
of that, assuming that, after passing each interval $\Delta z=1$, fields $%
u(\tau )$ and $v(\tau )$ are multiplied by a common factor $G=\exp \left(
\alpha \Delta z\right) $, whose value is selected so as to compensate the
loss in Eqs. (\ref{au}) and (\ref{av}). It is seen from the figure that the
periodic compensation may readily maintain the soliton's tailed shape.

\section{Stabilization of solitons by means of the group-velocity-mismatch
management}

\begin{figure}[tbp]
\subfigure[]{\includegraphics[width=3in]{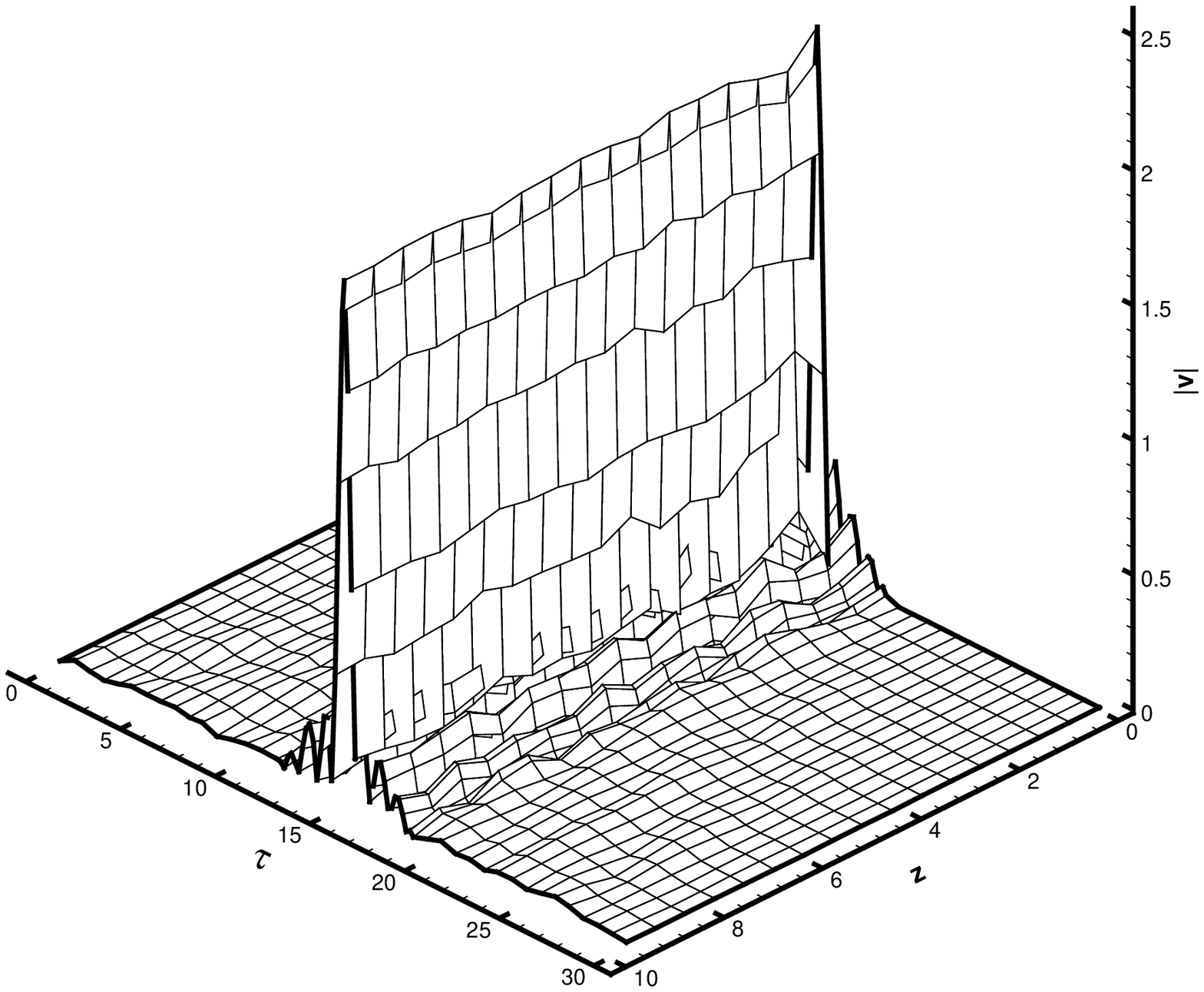}}\subfigure[]{%
\includegraphics[width=3in]{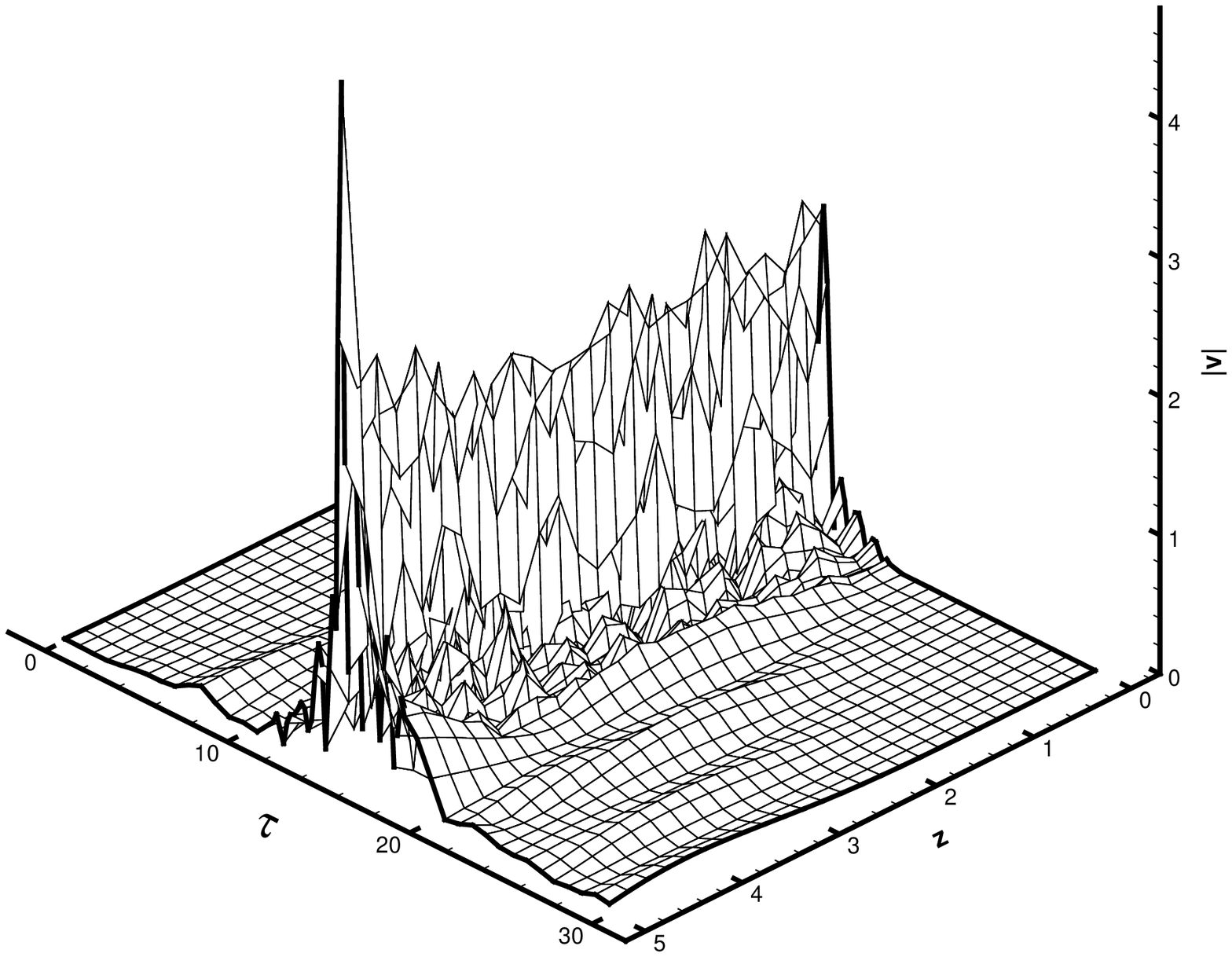}}
\caption{Evolution of solitons in the presence of the
group-velocity-mismatch term, with $c=-0.144$ (a) or $c=-1$ (b).
Other parameters are $D_{2}=0.37$ and $E=5,~R\equiv
K_{v}/K_{u}=1.13$.} \label{fig10}
\end{figure}

In this work, we do not aim to find numerically exact complex (\textit{%
chirped}) stationary solutions that may exist in the presence of the GVM, $%
c\neq 0$ (such solutions are known in the $\chi ^{(2)}$ model with the GVM
between the fundamental and second harmonics, which is related to the
existence of ``walking" $\chi ^{(2)}$ solitons \cite{walking}%
, and also in the system of coupled cubic NLS equations describing the
co-propagation of orthogonal linearly-polarized modes in optical fibers,
that includes, in addition to the XPM interaction, the nonlinear coupling
via the four-wave mixing \cite{Torner}). Actually, it would be quite
difficult to prepare such a pulse with a specific distribution of the
internal chirp for the use in the experiment. On the other hand, an
experimentally relevant problem is to simulate the evolution of the
previously found stationary solitons in the framework of Eqs. (\ref{u}) and (%
\ref{v}) that include the GVM term. To this end, in Fig.
\ref{fig10}(a) we display an example of the evolution of the soliton
in the presence of relatively weak GVM (the particular value,
$c=-0.144$, used in this figure, corresponds to physically relevant
parameters as per the above-mentioned lossless Drude model
\cite{Zhou}). It is seen that, while generating a permanent
perturbation of the soliton, the moderate GVM term does not destroy
it. On the other hand, Fig. \ref{fig10}(b) demonstrates that
stronger GVM, with $c=1$, gives rise to a very strong perturbation,
under which the soliton cannot maintain its integrity, even in an
approximate form.

\begin{figure}[tbp]
\includegraphics[width=4in]{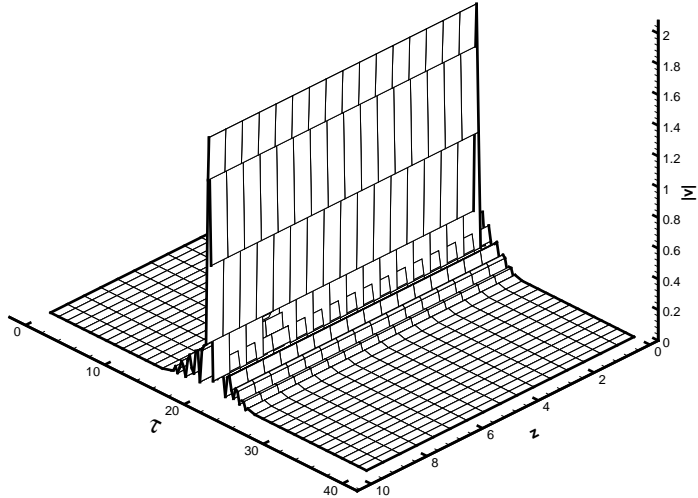}
\caption{The stabilization of the same soliton as in Fig.
\protect\ref{fig10}
provided by the GVM-management scheme based on map (\protect\ref{map}) with $%
L=0.2=0.25\cdot \left( 2\protect\pi /K_{v}\right) $.} \label{fig11}
\end{figure}

Because MMs and PCs are artificially built media, it may be quite
natural to consider possibilities to engineer superstructures in
them that can implement the GVM management, i.e., periodic
alternation of layers with large GVM coefficients of opposite signs.
In this way, it turns out easy to stabilize solitons against the
locally strong GVM, provided that the management period, $L$, is
relatively small in comparison with the effective wavelength,
$\lambda _{v}\equiv 2\pi /K_{v}$. A typical example of the
stabilization is displayed in Fig. \ref{fig11}. The scheme used in
this figure is based on the following \textit{management map}, with
the zero
average value of the GVM coefficient:%
\begin{equation}
c(z)=\left\{
\begin{array}{c}
-1,~0<z<L/4, \\
+1,~L/4<z<3L/4, \\
-1,~3L/4<z<L,%
\end{array}%
\right.  \label{map}
\end{equation}%
which repeats with period $L.$ In the case shown in Fig. \ref{fig11}, $L=0.2$%
, which amounts to $\lambda _{v}/4$. \textit{\ }

Finally, it is relevant to mention that, except for the case of $D_{2}=1$ in
Eq. (\ref{v}), the system as a whole does not support the Galilean
invariance, therefore moving solitons may be essentially different from the
quiescent ones, that were considered above. However, the study of moving
localized solutions within the framework of Eqs. (\ref{u}) and (\ref{v}) is
actually tantamount to considering solitons under the action of the GVM
effect.

\section{Conclusion}

The objective of this work was to study solitons in a model based on
the system of two XPM-coupled NLS equations, which describes the
co-propagation of two waves in MMs (metamaterials), if the loss may
be neglected, and in PCs (photonic crystals). The same system is
relevant for ordinary optical fibers, close to the zero-dispersion
point. The main peculiarity of the system is the small positive or
negative value of the GVD coefficient in one equation, while the
dispersion is fixed to be anomalous in the other. Unlike previously
studied systems of nonlinearly coupled NLS equations with equal GVD
coefficients, which gives rise only to simple stable single-peak
solitons with an arbitrary ratio of the energy between the
components, the present model generates soliton families which
feature complex shapes, including those with weakly localized
oscillatory tails and a double-peak maximum. Existence regions for
the single- and double-peak solitons have been identified in the
parameter planes, demonstrating a broad range of the bistability in
the system. Above the upper existence border, the two-component
solitons degenerate into obvious single-component fundamental NLS
solitons. Systematic direct simulations demonstrate that solitons of
both types are stable throughout their existence regions. Under the
action of the loss, the decaying solitons tend to evolve towards
more ``primitive" shapes, losing their tails. On the other hand, the
periodic compensation of the loss helps to maintain the original
shape. The numerical calculations were performed by means of a
properly adjusted pseudospectral method (while the usual split-step
algorithm may be problematic when applied to simulating the
evolution of solitons with conspicuous tails). The extended model,
including the GVM (group-velocity mismatch) was studied too. It has
been demonstrated that, while a sufficiently strong GVM term tends
to destroy the bimodal solitons, they can be readily stabilized by
way of the GVM-management scheme.

This work suggests extensions in several directions. First, a study of
collisions between moving solitons may be a natural addition to the above
analysis. Dispersion management \cite{my-book} may also be considered in the
framework of the present model. On the other hand, the model can be made
two- and three-dimensional by adding transverse coordinates. In that case, a
challenging issue would be a possibility to stabilize the respective
spatiotemporal solitons (``light bullets") against the collapse by means of
a grating created in the transverse direction(s), as well as by means of the
respectively modified dispersion-management scheme, cf. Ref. \cite{Warsaw}.


\begin{thebibliography}{99}
\bibitem{art} T. F. Krauss, R. M. De La Rue, S. Brand, Nature \textbf{383}
(1996) 699;

J. E. G. J. Wijnhoven, W. L. Vos, Science \textbf{281} (1998) 802;

C. P. Collier, T. Vossmeyer, J. R. Heath, Ann. Rev. Phys. Chem. \textbf{49}
(1998) 371;

R. J. Warburton, C. Schaflein, D. Haft D, F. Bickel, A. Lorke, K. Karrai, J.
M. Garcia, W. Schoenfeld, P. M. Petroff, Nature \textbf{405} (2000) 926;

S. Link, M. A. Ei-Sayed, Ann. Rev. Phys. Chem. \textbf{54} (2003) 331;

A. N. Grigorenko, A. K. Geim, H. F. Gleeson, Y. Zhang, A. A. Firsov, I. Y.
Khrushchev, J. Petrovic, Nature \textbf{438} (2005) 335.

\bibitem{Pendry} J. B. Pendry, D. R. Smith,
%Reversing light with negative refraction
Phys. Today \textbf{57} (2004) 37.%-43

\bibitem{superlense} J. B. Pendry,
%Negative refraction makes a perfect lens
Phys. Rev. Lett. \textbf{85} (2000) 3966;

N. Fang, H. Lee, C. Sun, X. Zhang,
%Sub-diffraction-limited optical imaging with a silver superlens
Science \textbf{308} (2005) 534;%-537

Z. Jacob, L. V. Alekseyev, E. Narimanov
%Optical hyperlens: Far-field imaging beyond the diffraction limit
, Opt. Exp. \textbf{14} (2006) 8247.%-8256

\bibitem{Shalaev} V. M. Shalaev, %Optical negative-index metamaterials
Nature Photonics \textbf{1} (2007) 41.%-48

\bibitem{Veselago} V. G. Veselago, Sov. Phys. Usp. \textbf{10} (1968) 509.

\bibitem{theory} J. B. Pendry, A. J. Holden, D. J. Robbins, W. J. Stewart,
%Magnetism from conductors and enhanced nonlinearity phenomena
IEEE Trans. Microwave Theory Tech. \textbf{47} (1999) 2075;

S. O. O'Brien, D. McPeake, S. A. Ramakrishna, J. B. Pendry,
%Near-infrared photonic band gaps and nonlinear effects in
%negative magnetic metamaterials
Phys. Rev. B \textbf{69} (2004) 241101(R);

A. A. Zharov, I. V. Shadrivov, Y. S. Kivshar,
%Nonlinear properties of left-handed metamaterials
Phys. Rev. Lett. \textbf{91} (2003) 03401;

I. R. Gabitov, R. A. Indik, N. M. Litchinitser, A. I. Maimistov, V. M.
Shalaev, J. E. Soneson,
%Double-resonant optical materials with embedded metal nanostructures
J. Opt. Soc. Am. \textbf{23} (2006) 535;%-542

N. M. Litchinitser, I. R. Gabitov, A. I. Maimistov, V. M. Shalaev,
%Effect of an optical negative index thin film on optical bistability
Opt. Lett. \textbf{32} (2007) 151;%-153

N. M. Litchinitser , I. R. Gabitov, A. I. Maimistov,
%Optical bistability in a nonlinear optical coupler with a negative index channel
Phys. Rev. Lett. \textbf{99} (2007) 113902.

\bibitem{SpecialTopics} A. I. Maimistov and I. R. Gabitov, Eur. Phys. J.
Special Topics \textbf{147} (2007) 265.

\bibitem{no-loss} %Generalized Nonlinear Schro¨dinger Equation for
%Dispersive Susceptibility and Permeability: Application to Negative
%Index Materials
M. Scalora, M. S. Syrchin, N. Akozbek, E. Y. Poliakov, G. D'Aguanno, N.
Mattiucci, M. J. Bloemer, A. M. Zheltikov, Phys. Rev. Lett. \textbf{95}
(2005) 013902;

P. P. Banerjee, G. Nehmetallah,
%Linear and nonlinear propagation in negative index materials
J. Opt. Soc. Am. B \textbf{23} (2006) 2348;

P. P. Banerjee, G. Nehmetallah, %Spatial and spatiotemporal solitary
%waves and their stabilization in nonlinear negative index materials
J. Opt. Soc. Am. B \textbf{24} (2007) A69.

\bibitem{Zhou} W. Zhou, W. Su, X. Cheng, Y. Xiang, X. Dai, S. Wen,
%Copropagation of two pulses of different frequencies and modulation
%instabilities induced by cross-phase modulation in metamaterials
Opt. Commun. \textbf{282} (2009) 1440.%-1447

\bibitem{Scalora} M. Scalora, G. D'Aguanno, M. Bloemer, M. Centini, D. de
Ceglia, N. Mattiucci, Y. S. Kivshar,
%Dynamics of short pulses and phase matched second harmonic generation in
%negative index materials
Opt. Exp. \textbf{14} (2006).

\bibitem{Haus} V. Roppo, M. Centini, C. Sibilia, M. Bertolotti, D. de
Ceglia, M. Scalora, N. Akozbek, M. J. Bloemer, J. W. Haus, O. G. Kosareva,
V. P. Kandidov,
%Role of phase matching in pulsed second-harmonic generation: Walk-off and
%phase-locked twin pulses in negative-index media
Phys. Rev. A \textbf{76} (2007) 033829.

\bibitem{solitons} A. D. Boardman, P. Egan, L. Velasco, N. King,
%Control of planar nonlinear guided waves and spatial solitons
%with a left-handed medium
J. Opt. A: Pure Appl. Opt. \textbf{7} (2005) S57.%-S67

I. V. Shadrivov, Y. S. Kivshar,
%Spatial solitons in nonlinear left-handed metamaterials
J. Opt. A: Pure Appl. Opt. \textbf{7} (2005) S68;%-S72

N. A. Zharova, I. V. Shadrivov, A. A. Zharov,
%Nonlinear transmission and spatiotemporal solitons in
%metamaterials with negative refraction
Opt. Exp. \textbf{13} (2005) 129;%-1298

M. Marklund, P. K. Shukla, L. Stenflo, G. Brodin,
%Solitons and decoherence in left-handed metamaterials
Phys. Lett. A \textbf{341} (2005) 231;%--234

S. C. Wen, Y. J. Xiang, W. H. Su, Y. H. Hu, X. Q. Fu, D. Y. Fan,
%Role of the anomalous self-steepening effect in modulation
%instability in negative-index material
Opt. Exp. \textbf{14} (2006) 1568;%-1575

Y. M. Liu, G. Bartal, D. A. Genov, X. Zhang,
%Subwavelength discrete solitons in nonlinear metamaterials
Phys. Rev. Lett. \textbf{99} (2007) 153901;

I. Kourakis, N. Lazarides, G. P. Tsironis,
%Self-focusing and envelope pulse generation in nonlinear magnetic metamaterials
Phys. Rev. E \textbf{75} (2007) 067601.

\bibitem{Agrawal} G. P. Agrawal,
%Modulation instability induced by cross-phase modulation
Phys. Rev. Lett. \textbf{59} (1987) 880.%-883

G. P. Agrawal, P. L. Baldeck, R. R. Alfano,
%Modulation instability induced by cross-phase modulation in optical fibers
Phys. Rev. A \textbf{39} (1989) 3406.%-3413

\bibitem{DW} B. A. Malomed%Optical domain walls
, Phys. Rev. E \textbf{50} (1994) 1565.%-1571

\bibitem{PCF} A. Ferrando, E. Silvestre, P. Andr\'{e}s, J. J. Miret, M. V.
Andr\'{e}s,
%Designing the properties of dispersion-flattened photonic crystal fibers
Opt. Exp. \textbf{9} (2001) 687;%-697

T. M. Monro, D. J. Richardson,
%Holey optical fibres: Fundamental properties and device applicationsCompt
Comptes Rendus Phys. \textbf{4}(2003) 175.%-186

\bibitem{book} J. D. Joannopoulos, S. G. Johnson, J. N. Winn, R. D. Meade,
\textit{Photonic crystals: molding the flow of light} (Princeton University
Press: Princeton and Oxford, 2008).

\bibitem{loss} R. W. Ziolkowski, E. Heyman,
%Wave propagation in media having negative permittivity and permeability
Phys. Rev. E \textbf{64} (2001) 056625;

A. Alu, N. Engheta, Phys. Rev. E \textbf{72} (2005) 016623;

G. Dolling, C. Enkrich, M. Wegener, C. M. Soukoulis, S. Linden,
%Low-loss negative-index metamaterial at telecommunication wavelengths
Opt. Lett. \textbf{31} (2006) 1800;%-1802

N. M. Litchinitser, V. M. Shalaev,
%Metamaterials Loss as a route to transparency
Nature Photonics \textbf{3} (2009) 75.%-76

\bibitem{necessary-loss} M. I. Stockman,
%Criterion for negative refraction with low optical losses from
%a fundamental principle of causality
Phys. Rev. Lett. \textbf{98} (2007) 177404.

\bibitem{matched-impedance} V. M. Shalaev, W. Cai, U. K. Chettiar, H. Yuan,
A. K. Sarychev, V. P. Drachev, A. V. Kildishev, Opt. Lett. \textbf{30}
(2005) 3356.

\bibitem{gain} A. K. Popov, V. M. Shalaev, Opt. Lett. \textbf{31} (2006)
2169;

A. D. Boardman, Yu. G. Rapoport, N. King, V. N. Malnev,
%Creating stable gain in active metamaterials
J. Opt. Soc. Am. B \textbf{24} (2007) A53;%-A61

J. A. Gordon, R. W. Ziolkowski,
%The design and simulated performance of a coated nano-particle laser
Opt. Exp. \textbf{15} (2007) 2622.%-2653

A. K. Sarychev, G. Tartakovsky,
%Magnetic plasmonic metamaterials in actively pumped host medium and plasmonic nanolaser
Phys. Rev. B \textbf{75} (2007) 085436;

M. A. Noginov, V. A. Podolskiy, G. Zhu, M. Mayy, M. Bahoura, J. A. Adegoke,
B. A. Ritzo, K. Reynolds,
%Compensation of loss in propagating surface plasmon polariton by
%gain in adjacent dielectric medium
Opt. Exp. 16 (2008) 1385.%-1392

\bibitem{cavity} P. Tassin, L. Gelens, J. Danckaert, I. Veretennicoff , G.
Van der Sande, P. Kockaert, M. Tlidi,
%Dissipative structures in left-handed material cavity optics
Chaos \textbf{17} (2007) 037116.

\bibitem{Ship} A. Shipulin, G. Onishchukov, B. A. Malomed, J. Opt. Soc. Am.
B \textbf{14} (1997) 3393;%-3402

B. A. Malomed A. Shipulin,
%Stabilization of optical pulse propagation by a supporting periodic structure
Opt. Commun. \textbf{162} (1999) 140.%-147

\bibitem{numerical} D. N. Christodoulides and R. I. Joseph,
%VECTOR SOLITONS IN BIREFRINGENT NONLINEAR DISPERSIVE MEDIA
Opt. Lett. \textbf{13} (1988) 53;

M. Haelterman, A. P. Sheppard, A. W. Snyder,
%BOUND-VECTOR SOLITARY WAVES IN ISOTROPIC NONLINEAR DISPERSIVE MEDIA
Opt. Lett. \textbf{18} (1993) 1406.%-1408

\bibitem{variational} T. Ueda, W. L. Kath,
%DYNAMICS OF COUPLED SOLITONS IN NONLINEAR OPTICAL FIBERS
Phys. Rev. A \textbf{42} (1990) 563;%-571

D. J. Kaup, B. A. Malomed, R. S. Tasgal,
% INTERNAL DYNAMICS OF A VECTOR SOLITON IN A NONLINEAR-OPTICAL FIBER
Phys. Rev. E \textbf{48} (1993) 3049;%-3053

B. A. Malomed, R. S. Tasgal,
% Internal vibrations of a vector soliton in the coupled nonlinear Schrodinger equations
Phys. Rev. E \textbf{58} (1998) 2564.%-2575

\bibitem{Yang} J. Yang,
%Classification of the solitary waves in coupled nonlinear Schroedinger equations,
Physica D \textbf{108} (1997) 92.%-112

\bibitem{aboutMoti2} D. E. Pelinovsky, J. Yang,
%Instabilities of Multihump Vector Solitons in Coupled
%Nonlinear Schroedinger Equations
Stud. Appl. Math. \textbf{115} (2005) 109.%-137

\bibitem{Moti1} M. Mitchell, M. Segev, T. H. Coskun, D. N. Christodoulides,
%Theory of Self-Trapped Spatially Incoherent Light Beams
Phys. Rev. Lett. \textbf{79} (1997) 4990.

\bibitem{Moti2} M. Mitchell, M. Segev, D. N. Christodoulides,
%Observation of Multihump Multimode Solitons
Phys. Rev. Lett. \textbf{80} (1998) 4657.

\bibitem{aboutMoti1} E. A. Ostrovskaya, Y. S. Kivshar, D. V. Skryabin, W. J.
Firth, %Stability of multihump optical solitons
Phys. Rev. Lett. \textbf{83} (1999) 296.%-299

\bibitem{Moti2D} T. Carmon, C. Anastassiou, S. Lan, D. Kip, Z. H.
Musslimani, M. Segev, D. Christodoulides,
%Observation of two-dimensional multimode solitons
Opt. Lett. \textbf{25} (2000) 1113;%-1115

Z. H. Musslimani, M. Segev, D. N. Christodoulides, M. Soljacic,
%Composite multihump vector solitons carrying topological charge
Phys. Rev. Lett. \textbf{84} (2000) 1164.%-1167

\bibitem{chi2} C. R. Menyuk, R. Schiek, L. Torner,
%SOLITARY WAVES DUE TO CHI((2))/CHI((2)) CASCADING
J. Opt. Soc. Am. B \textbf{11} (1994) 2434;%-2443

P. Di Trapani, D. Caironi, G. Valiulis, A. Dubietis, R. Danielius, A.
Piskarskas,
%Observation of temporal solitons in second-harmonic generation with tilted pulses
Phys. Rev. Lett. \textbf{81} (1998) 570;%-573

X. Liu, L. J. Qian, F. W. Wise,
%Generation of optical spatiotemporal solitons
Phys. Rev. Lett. \textbf{82} (1999) 463.%-4634

\bibitem{Kale} K. Beckwitt, Y. F. Chen, F. W. Wise, B. A. Malomed,
%Temporal solitons in quadratic nonlinear media with opposite
%group-velocity dispersions at the fundamental and second harmonics
Phys. Rev. E \textbf{68} (2003) 057601.

\bibitem{we1} P. Y. P. Chen, B. A. Malomed,
%Stability of temporal solitons in uniform and "managed" quadratic
%nonlinear media with opposite group-velocity dispersions at
%fundamental and second harmonics
Opt. Commun. \textbf{281} (2008) 5257;%-5266

P. Y. P. Chen, B.A. Malomed, \textit{Stabilization of spatiotemporal
solitons in second-harmonic-generating media}, Opt. Commun., in press.

\bibitem{tandem} L. Torner,
%Walkoff-compensated dispersion-mapped quadratic solitons
IEEE Phot. Tech. Lett. \textbf{11} (1999) 1268;%
%-1270

L. Torner, S. Carrasco, J. P. Torres, L. C. Crasovan, D. Mihalache,
%Tandem light bullets
Opt. Commun. \textbf{199} (2001) 277;%-281

S. Carrasco, D. V. Petrov, J. P. Torres, L. Torner, H. Kim, G. Stegeman, J.
J. Zondy,
%Observation of self-trapping of light in walk-off-compensating tandems
Opt. Lett. \textbf{29} (2004) 382.%-384

\bibitem{pseudo} D. Gottlieb, S. A. Orszag, \textit{Numerical Analysis of
Spectral Method: Theory and Applications} (SIAM: Philadelphia, 1977);

G. Cohen, S. Fanqueux,
%Mixed spectral finite elements for the linear elsticity system in unbounded domains
SIAM J. Sci. Comput. \textbf{26} (2005) 864.%-884

\bibitem{pseudo-soliton} M. Dehghan, A. Taleei,
%Numerical Solution of Nonlinear Schroedinger Equation by Using
%Time-Space Pseudo-Spectral Method
Numerical Methods for Partial Differential Equations, in press (2009; DOI
10.1002/num.20468).

\bibitem{split-step} O. V. Sinkin, R. Holzlohner, J. Zweck, C. R. Menyuk,
%Optimization of the split-step Fourier method in modeling
%optical-fiber communications systems
J. Lightwave Tech. 21 (2003) 61.%-68

\bibitem{Isaac} I. N. Towers, B. A. Malomed, F. W. Wise,
%Light bullets in quadratic media with normal dispersion at the second harmonic
Phys. Rev. Lett. \textbf{90} (2003) 123902.

\bibitem{Arik} A. Gubeskys, B. A. Malomed, I. M. Merhasin,
%Alternate solitons: Nonlinearly-managed one- and two-dimensional
%solitons in optical lattices
Stud. Appl. Math. \textbf{115} (2005) 255%-277
; H. Sakaguchi and B. A. Malomed,
%Solitary vortices and gap solitons in rotating optical lattices.
Phys. Rev. A \textbf{79} (2009) 043606.

\bibitem{Sadhan} S. K. Adhikari, B. A. Malomed, Phys. Rev. A \textbf{77}
(2008) 023607.

\bibitem{walking} L. Torner, D. Mazilu, D. Mihalache,
%Walking solitons in quadratic nonlinear media
Phys. Rev. Lett. \textbf{77} (1996) 2455;%-2458

U. Peschel, F. Lederer, and B. A. Malomed,
%Stability of temporal chirped solitary nonlinear optical waves
%in quadratically nonlinear media
Phys. Rev. E \textbf{55} (1997) 6155.

\bibitem{Torner} D. Mihalache, D. Mazilu, L. Torner,
%Stability of Walking Vector Solitons
Phys. Rev. Lett. \textbf{81} (1998) 4353.

\bibitem{my-book} B. A. Malomed, \textit{Soliton management in Periodic
Systems} (Springer: New York, 2006).

\bibitem{Warsaw} M. Matuszewski, M. Trippenbach, B. A. Malomed, E. Infeld,
M. Skorupski,
%Two-dimensional dispersion-managed light bullets in Kerr media
Phys. Rev. E \textbf{70} (2004) 016603.
\end{thebibliography}
\end{document}